\documentclass[a4paper,11pt]{article}
\pdfoutput=1 

\usepackage{jcappub} 

\usepackage[T1]{fontenc} 

\usepackage{soul}

\usepackage{amsmath}
\usepackage{commath}
\usepackage{mleftright}
\usepackage{bm}

\usepackage{subcaption}
\usepackage{wrapfig}
\usepackage{graphicx} 
\graphicspath{{figures/}} 
\usepackage{cprotect}

\usepackage{xcolor}

\newcommand{\polarization}{}

\newcommand{\map}[1]{\bm{#1}}
\newcommand{\myvector}[1]{\mathbf{#1}}
\newcommand{\mymatrix}[1]{\mathbf{#1}}
\newcommand{\maparray}[1]{\bm{#1}}
\newcommand{\myset}[1]{{#1}}
\newcommand{\experiment}[1]{\verb|#1|}

\makeatletter
\newcommand\notsotiny{\@setfontsize\notsotiny\@vipt\@viipt}
\makeatother

\title{On the Detection of CMB B-modes from Ground at Low Frequency}

\author[a,b]{E. de la Hoz,}
\author[a]{P. Vielva,}
\author[a]{R. B. Barreiro}
\author[a]{and E. Mart\'inez-Gonz\'alez}

\affiliation[a]{Instituto de F\'isica de Cantabria, CSIC-Universidad de Cantabria,\\ 
Avda. de los Castros s/n, E-39005 Santander, Spain}
\affiliation[b]{Dpto. de Física Moderna, Universidad de Cantabria, \\
Avda. los Castros s/n, E-39005 Santander, Spain}

\emailAdd{delahoz@ifca.unican.es}
\emailAdd{vielva@ifca.unican.es}
\emailAdd{barreiro@ifca.unican.es}
\emailAdd{martinez@ifca.unican.es}

\abstract{The primordial CMB $B$-mode search is on the spotlight of the scientific community due to the large amount of cosmological information that is encoded in the primeval signal. However, the detection of this signal is challenging from the data analysis point of view, due to the relative low amplitude compared to the foregrounds, the lensing contamination coming from the leakage of $E$-modes, and the instrumental noise. Here, we studied the viability of the detection of the primordial polarization $B$-mode with a ground-based telescope operating in the microwave low-frequency regime (i.e., from 10GHz-120GHz) in a handful of different scenarios: i. the instrument's channels distribution and noise, ii. the tensor-to-scalar ratio ($r$) detectability considering different possible $r$ values and degrees of delensing,  iii. the effect of including a possible source of polarized anomalous microwave emission (AME), iv. the strengths and weaknesses of different observational strategies and, v. the atmospheric and systematic noise impact on the recovery. We focused mainly on the removal of galactic foregrounds as well as noise contamination by applying a full-parametric pixel-based maximum likelihood component separation technique. Moreover, we developed a numerical methodology to estimate the residuals power spectrum left after component separation, which allow us to mitigate possible biases introduced in the primordial $B$-mode power spectrum reconstruction. Among many other results, we found that this sort of experiment is capable of detecting Starobinsky's $r$ even when no delensing is performed or, a possible polarized AME contribution is taken into account. Besides, we showed that this experiment is a powerful complement to other on-ground or satellite missions, such as LiteBIRD, since it can help significantly with the low-frequency foregrounds characterization. }

\begin{document}
\maketitle
\flushbottom

\section{Introduction}
\label{sec:introduction}

For several decades the scientific community  has devoted a tremendous effort towards the improvement in the Cosmic Microwave Background (CMB) polarization detection. The interest arises due to the large amount of cosmological information comprised in it, e.g., the predicted primordial $B$-modes.  Primordial $B$-modes are only sourced by non-scalar perturbations, hence a detection would constitute a definitive proof of the existence of primordial gravitational waves (PGWs) \cite{guth1981inflationary,linde1982new,starobinsky1982dynamics}. Even though PGWs are conjectured by most of inflationary models, their predictions differ in the PGWs' amplitude. Current constraints on the tensor-to-scalar perturbations ratio $r$ are $\lesssim 0.056$ at 95 \% CL \cite{akrami2018planck_inflation}, which reveal the faintness of this signal. Unfortunately, there is no theoretical lower bound for this quantity so there is no warranty of detection. However, even in a non-detection case, all these endeavors would not be futile as more sensitive instruments can place stronger constraints in the PGWs' amplitude and debunk a considerable number of inflationary models \cite{akrami2018planck_inflation,martin2014best}. \\
\\
Currently, there are many planned ground-based experiments, e.g., CMB-S4 \cite{abazajian2016cmb}, Simons Observatory \cite{ade2019simons}, BICEP array \cite{hui2018bicep}, as well as satellite missions, e.g., LiteBIRD \cite{matsumura2016litebird}, PICO \cite{hanany2019pico}, which include the primordial $B$-mode search among their top scientific goals. Their primary objective is to be able to detect, or at least constrain $r$ with a sensitivity $\sigma_r (r=0) \leq 10^{-3}$. This work constitutes a preliminary study of the performance of a potential on-ground experiment encompassed in this international $B$-mode chase. This experiment is proposed in the context of the European Low Frequency Survey initiative. Here, we have studied a ground-based instrument to perform a Low Frequency Survey (LFS) with the following characteristics: operation in the low-frequency range covering approximately 10-120 GHz, full-sky coverage, and finally, capability of placing stringent constraints on $r$.  \\
\\
The main problem that the search of these primordial modes faces is the signal's weakness. Moreover, this elusive signal hides among other $B$-mode sources with rather different origins such as:  $E$-modes converted to $B$-modes due to gravitational lensing along the photons path, foreground contaminants like the synchrotron or the thermal dust emissions, instrumental noise, etc. Therefore, special data treatment methods are required to disentangle the primeval signal from the nuisance signals. Here, we have focused mainly on component separation methods which deal with the foreground and noise contamination \cite{ichiki2014cmb}. To conduct the instrument's forecasts we have applied a full-parametric pixel-based maximum-likelihood component separation method. Furthermore, we have developed an approach to estimate a model of the foregrounds and noise residuals. A residuals model allow us to correct possible biases induced in cosmological parameters due to insufficient foreground removal \cite{errard2019characterizing}, as well as to forecast which values of $r$ are detectable. 
\\
\\
We have applied the aforementioned method to study the instruments performance in different situations: i. the instrument's channels distribution and noise, to determine the most optimal setup that fulfills the $\sigma_r$ constraint,  ii. the $r$ detectability considering different possible $r$ values and degrees of delensing, i.e., fraction of $E$-to-$B$ modes removed, iii. the effect of including a possible source of polarized anomalous microwave emission (AME), iv. the strengths and weaknesses of different observational strategies, e.g., full-sky vs. small sky patches observation in the same observational time and, v. the atmospheric/systematic noise impact on the recovery. 
\\
\\
In addition, this instrument is proposed to be a potential complement to other experiments, both on-ground and satellite. It has been shown \cite{stivoli2010maximum} that different frequency coverage can affect the level of foreground residuals. Since this experiment studies the  low-frequency regime with sensitivities never achieved before, it can help with the characterization of  foregrounds dominant in this range, e.g., the synchrotron, and AME. Here, the reconstruction of the foreground components is analyzed in the case of LiteBIRD alone, and LiteBIRD with this telescope. 
\\
\\
This work is structured as follows: in section~\ref{sec:observational_configuration} we describe the observational characteristics of this instrument; in section~\ref{sec:sky_model} the different contributions to the sky simulations, i.e., the astrophysical signals as well as the instrumental noise, are described; section~\ref{sec:component_separation} outlines the component separation approach followed; we explain the residuals model estimation methodology conducted in this work in section~\ref{sec:self-residual_estimation}; in section~\ref{sec:experiment_performance} we compare the telescope performance under different scenarios, i.e., different experimental setups, noise scaling, etc.;  section~\ref{sec:complementarity_LiteBIRD} studies the improvement on LiteBIRD's foreground characterization when combined with this experiment; finally, we draw some conclusions in section~\ref{sec:conclusions}.

\section{Observational Configuration}
\label{sec:observational_configuration}

In this section we highlight the basic observational characteristics of the propounded experiment, i.e., the instrument's location, the sky coverage, and the frequency range covered.

\paragraph{Experiment location:} This experiment is thought to be capable of measuring the whole sky. Therefore it requires at least two facilities, one located at the Northern Hemisphere (NH) and the other at the Southern Hemisphere (SH). A plausible choice for the NH location is Tenerife, in the Canary Islands, because there are precursor experiments successfully operating in the low-frequency regime like \verb|QUIJOTE| \cite{rubino2012quijote}. On the other hand, Atacama is another realistic option for the SH since a handful of CMB experiments are already settled there, e.g., \verb|ACTPol| \cite{niemack2010actpol}, \verb|ABS| \cite{essinger2009atacama}, \verb|CLASS| \cite{essinger2014class}, due to its sky quality.

\begin{figure}
    \centering
    \begin{subfigure}[t]{0.33\textwidth}
        \centering
        \includegraphics[width=\linewidth]{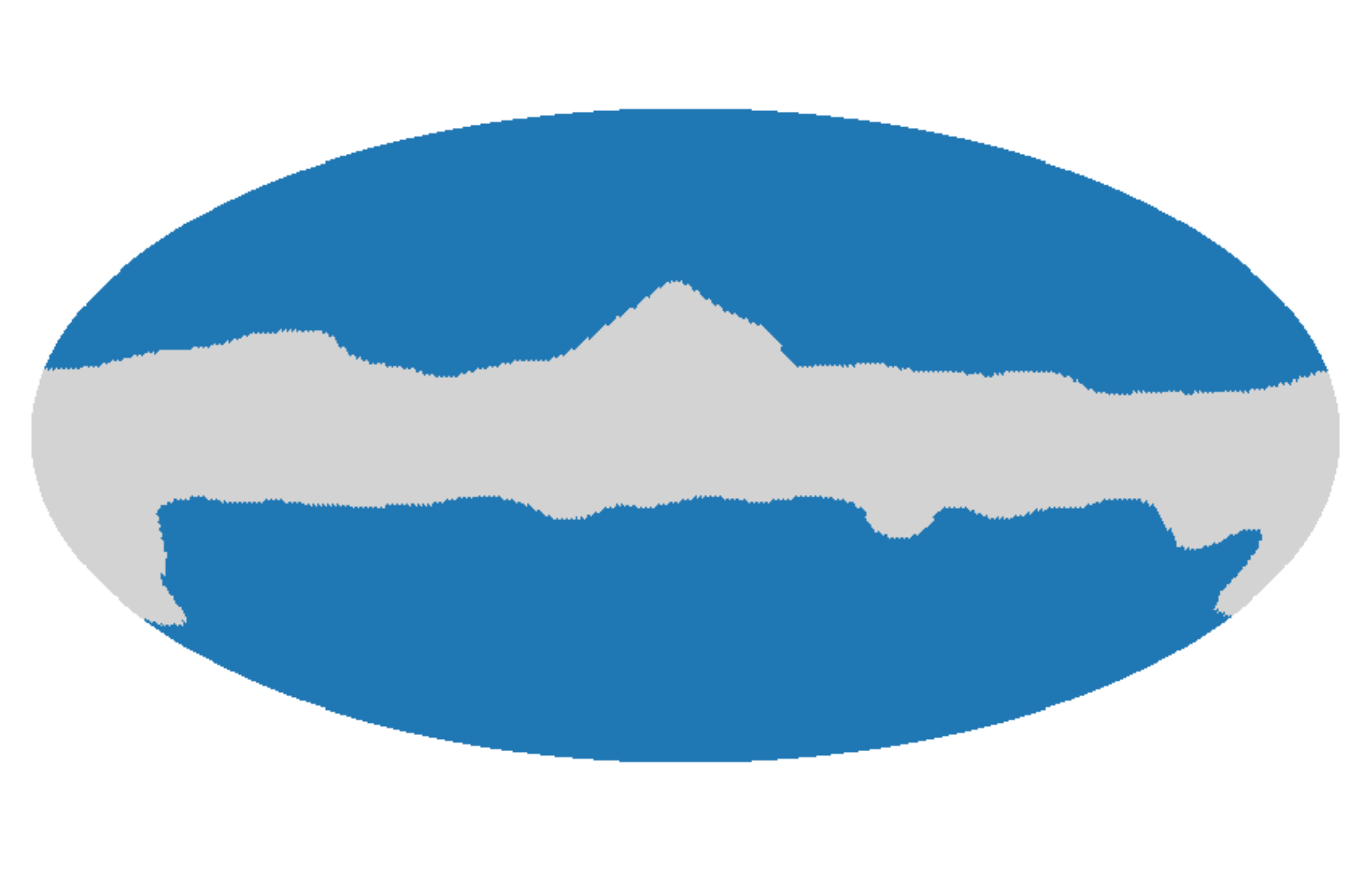}
        \cprotect\caption{\verb|Planck|: 70\% galactic plane mask}
        \label{subfig:planck_mask}
    \end{subfigure}%
    \begin{subfigure}[t]{0.33\textwidth}
        \centering
        \includegraphics[width=\linewidth]{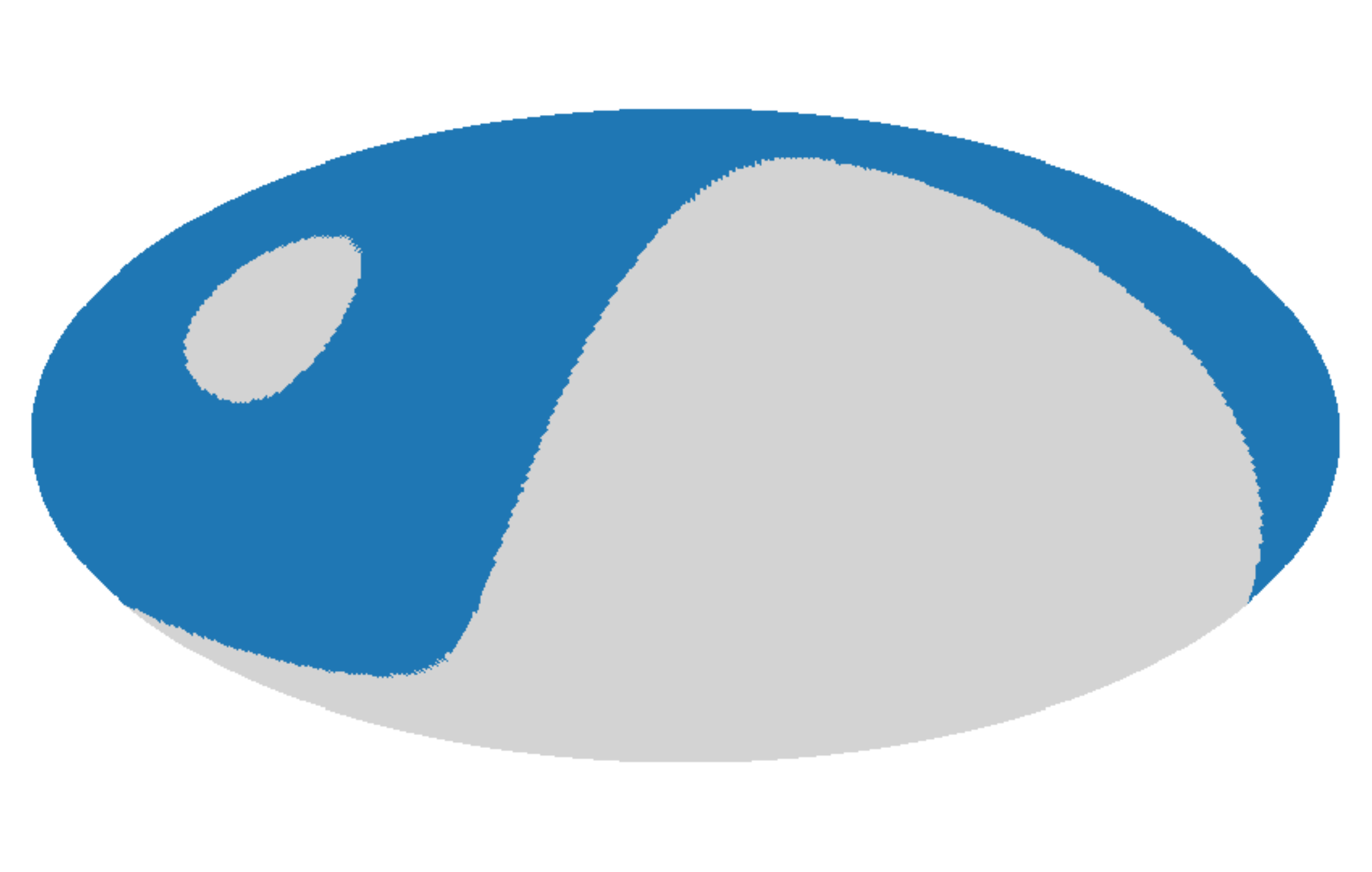}
        \cprotect\caption{\texttt{QUIJOTE}: Wide Survey}
        \label{subfig:quijote_masks}
    \end{subfigure}
    \begin{subfigure}[t]{0.33\textwidth}
        \centering
        \includegraphics[width=\linewidth]{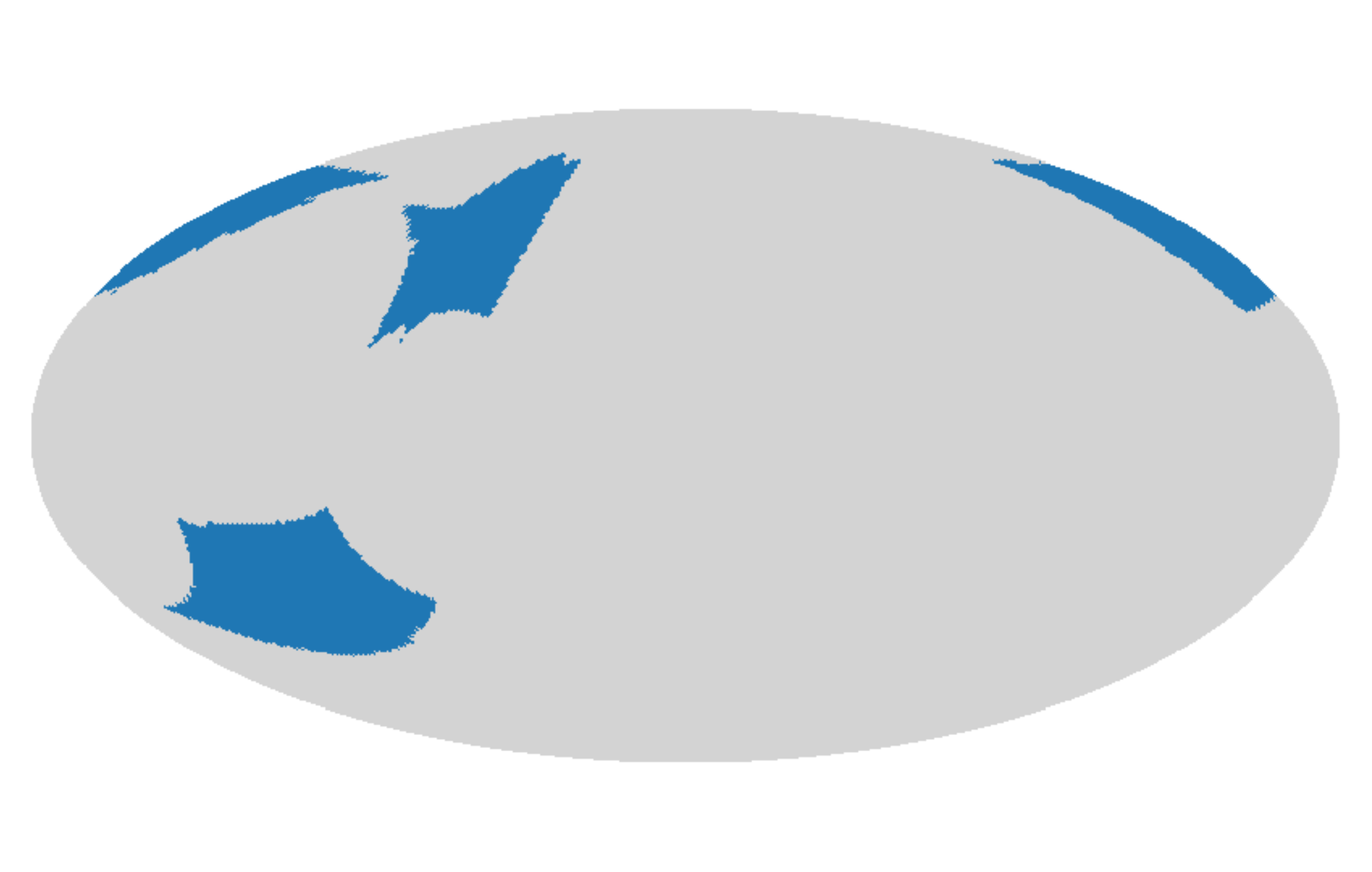}
        \cprotect\caption{\texttt{QUIIJOTE}: Cosmological-areas}
        \label{subfig:cosmoareas_masks}
    \end{subfigure}
    \caption{Sky observational masks.}
    \label{fig:masks}
\end{figure}

\paragraph{Sky coverage:} In this study we have applied three distinct observational masks in order to simulate different experiment locations and scanning strategies.
\begin{itemize}
    \item  In the case of two instruments located one at the NH and another at the SH, the full sky is available but we have applied a galactic disk mask to remove the foreground most contaminated areas. The mask is obtained from Planck Legacy Archive\footnote{	
The $f_{sky} = 0.7$ (fraction of available sky) galactic mask from \texttt{HFI\_Mask\_GalPlane-apo0\_2048\_R2.00.fits} downloaded from  \url{https://pla.esac.esa.int/##maps}} \cite{ade2014planck}. 
    \item Another option is when the instrument is located only at NH, e.g., in Tenerife, hence, only a fraction of the sky is available. Thus, in this work we have considered the same observable sky as \verb|QUIJOTE| \cite{poidevin2018awy}. Moreover, as in the previous situation we have applied a galactic mask to remove foreground dominated areas (same galactic area as in  the previous \verb|Planck| mask).
    \item Besides, instead of observing the whole accessible sky, one can observe small sky patches where the foregrounds are less dominant. An advantage of this strategy is the increase in the signal-to-noise ratio due to spending more time in a particular area within the same observational time. To study this option we have considered  the \verb|QUIJOTE|'s cosmological areas mask, i.e., sky patches where the foreground contamination is less harmful \cite{poidevin2018awy}.
\end{itemize}
The observational masks described are shown in figure~\ref{fig:masks}. In the NH case we apply a combined mask from figure~\ref{subfig:planck_mask} and figure~\ref{subfig:quijote_masks}.

\paragraph{Frequency range:}
The LFS is designed to operate in the low-frequency regime, i.e., from 10 to 120 GHz, with its channels distributed among three frequency bands, where the atmospheric absorption is less significant. The bands are listed below:
\begin{itemize}
    \item Low-frequency band (lb) from 10-20 GHz.
    \item Middle-frequency band (mb) from 26-46 GHz.
    \item High-frequency band (hb) from 75-120 GHz.
\end{itemize}
A telescope setup is defined by a 3-tuple [$n_{lb}$,$n_{mb}$,$n_{hb}$] where $n_{b}$ is the number of channels in the band $b$. The frequency channels within a band are distributed evenly as follows
\begin{equation}
    \nu_{i,b} = \nu_{ini,b} + \dfrac{\Delta \nu_{b}}{2n_{b}}(2i-1) \, ,
\end{equation}
where $\nu_{i,b}$ is the $i$-th central frequency of the $b$ band in a given experimental configuration, $\nu_{ini,b}$ is the lowest frequency within that band, and $\Delta \nu_{b}$ is the $b$ band's bandwidth. For example, if the experimental setup is [5,5,5], the frequency channels centers are (11, 13, 15, 17, 19), (28, 32, 36, 40, 44), and (79.5, 88.5, 97.5, 106.5, 115.5) GHz in the lb, mb, and hb respectively. 

\section{Sky Model}
\label{sec:sky_model}

Here, we describe the procedure adopted to generate the simulated maps used in the forecasts. Multi-frequency simulations are generated at a resolution of $n_{side} = 64$ (for observations of the whole available sky), and $n_{side} = 256$ (observations of small sky patches) using \texttt{HEALPix}\footnote{Hierarchical Equal Area isoLatitude Pixelization, \url{https://healpix.sourceforge.io/}, \cite{gorski2005healpix}.}. The frequencies selected depend on the telescope setup considered, and the effect of the detectors bandwidth is not taken into account, i.e., the channel's bandwidth is modeled as a  $\delta$-function. \\
\\
Our sky simulations contain the following components: CMB, galactic foregrounds\footnote{Contamination due to point sources emission is neglected since its effect is not significant at the resolutions studied.}, and other inevitable noise sources, such as instrumental or atmospheric/systematic noise. 
\begin{figure}
    \begin{minipage}{.58\linewidth}
    \centering
    \includegraphics[width=\linewidth]{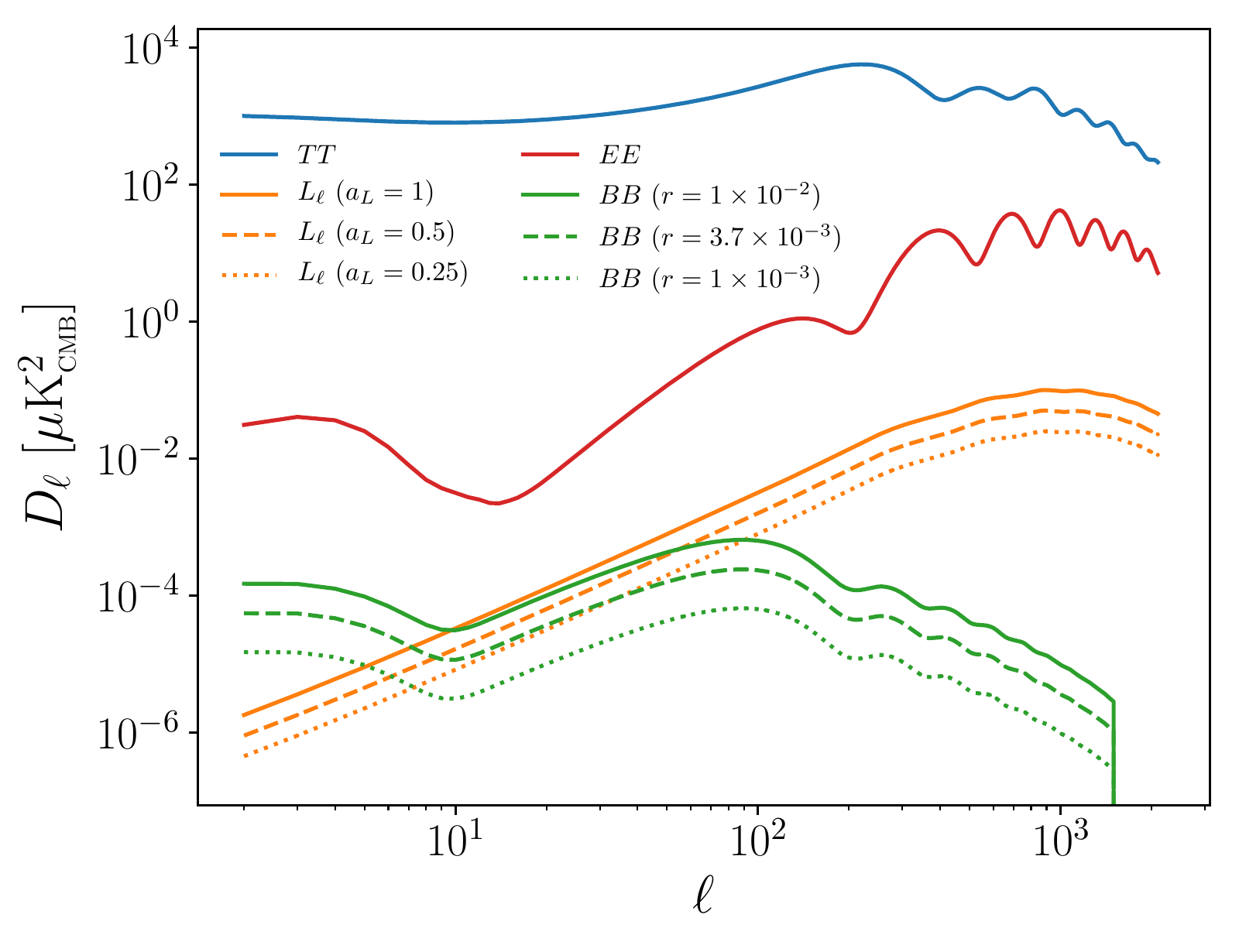}
    \captionof{figure}{CMB spectra. $TT$, $EE$, $BB$ and $E$-to-$B$ lensing ($L$) contributions are displayed.}
    \label{fig:dls_cmb}
    \end{minipage}
    \hspace{.02\linewidth}
    \begin{minipage}{.4\linewidth}
    \centering
    \begin{tabular}{|cc|}
        \hline
        $r \times 10^{3}$ & $a_{L}$ \\
        \hline
        0 & 1\\
        0 & 0.5\\
        3.7 & 1\\
        3.7 & 0.5\\
        \hline
    \end{tabular}
    \captionof{table}{\textbf{Cosmological values.} Combinations of $r$ and $a_L$ values used for the CMB simulations. The value $r = 3.7 \times 10^{-3}$ is the expected Starobinsky \cite{lyth1999particle} value according to the latest Planck results \cite{aghanim2018planck}.}
    \label{tab:CMB_maps_values}
    \end{minipage}
\end{figure}
\paragraph{CMB:} CMB maps are drawn as Gaussian random realizations of theoretical power spectra. The power spectra are evaluated with the Boltzmann-solver \texttt{CAMB} \cite{lewis2011camb} using the latest cosmological parameters from Planck \cite{aghanim2018planck}. Figure~\ref{fig:dls_cmb} shows the $D_{\ell}$\footnote{$D_{\ell} \equiv C_{\ell} \ell(\ell+1)/(2\pi)$, where $C_{\ell}$ stands for the angular power spectrum.} of the primordial $B$-mode for different $r$ values as well as the $E$-to-$B$ lensing contamination modes, the $EE$ and $TT$. With the lowest resolution, $n_{side} = 64$, multipoles as high as $\ell_{max} = 3n_{side} - 1 \sim 190$  are reached, hence both the re-ionization and recombination bumps could be observed. \\
\\
In this work we have considered different scenarios with simulated CMB maps whose $r$ and $a_L$ ($E$-to-$B$ lensing amplitude assuming a certain level of delensing) values, listed in table~\ref{tab:CMB_maps_values}, differ. To allow meaningful comparisons among those scenarios, CMB maps were generated from template $a_{\ell m}$ realizations. The procedure followed to generate the simulated CMB maps is explained below.
\begin{enumerate}
    \item Two sets of $a_{\ell m}$ were generated using the \texttt{synalm} routine of \texttt{healpy}, a python implementation of \texttt{HEALPix} \cite{zonca2019healpy}. One set $\{t^{ul}_{\ell m},e^{ul}_{\ell m},b^{ul}_{\ell m}\}$ is created from a collection of unlensed power spectra with $r=1$, and another $\{t^{l}_{\ell m},e^{l}_{\ell m},b^{l}_{\ell m}\}$ from lensed power spectra with $r=0$.
    \item Then, CMB maps were generated using the \texttt{healpy} routine \texttt{alm2map} using the following set of $a_{\ell m}$
    \begin{equation}
        t_{\ell m} = t^{l}_{\ell m} \, , 
        \qquad
        e_{\ell m} = e^{l}_{\ell m} \, ,
        \qquad
        b_{\ell m} = \sqrt{r}b^{ul}_{\ell m} + \sqrt{a_L}b^{l}_{\ell m} \, .
        \label{eq:alm2map_cmb}
    \end{equation}
    These maps were smoothed with a FWHM = 1$^{\circ}$ or $15$ arcmin for the $n_{side} = 64$, $256$ resolutions respectively, and corrected with the appropriate pixel window function term. Note that the $t_{\ell m}$ and $e_{\ell m}$ terms do not have the contribution from the tensor fluctuations. However, since the tensor-to-scalar ratio is small, the possible errors that may arise from this mismatch are almost negligible.
\end{enumerate}

\begin{figure}
    \centering
    \begin{minipage}[t]{.48\linewidth}
        \centering
        \includegraphics[width=\linewidth]{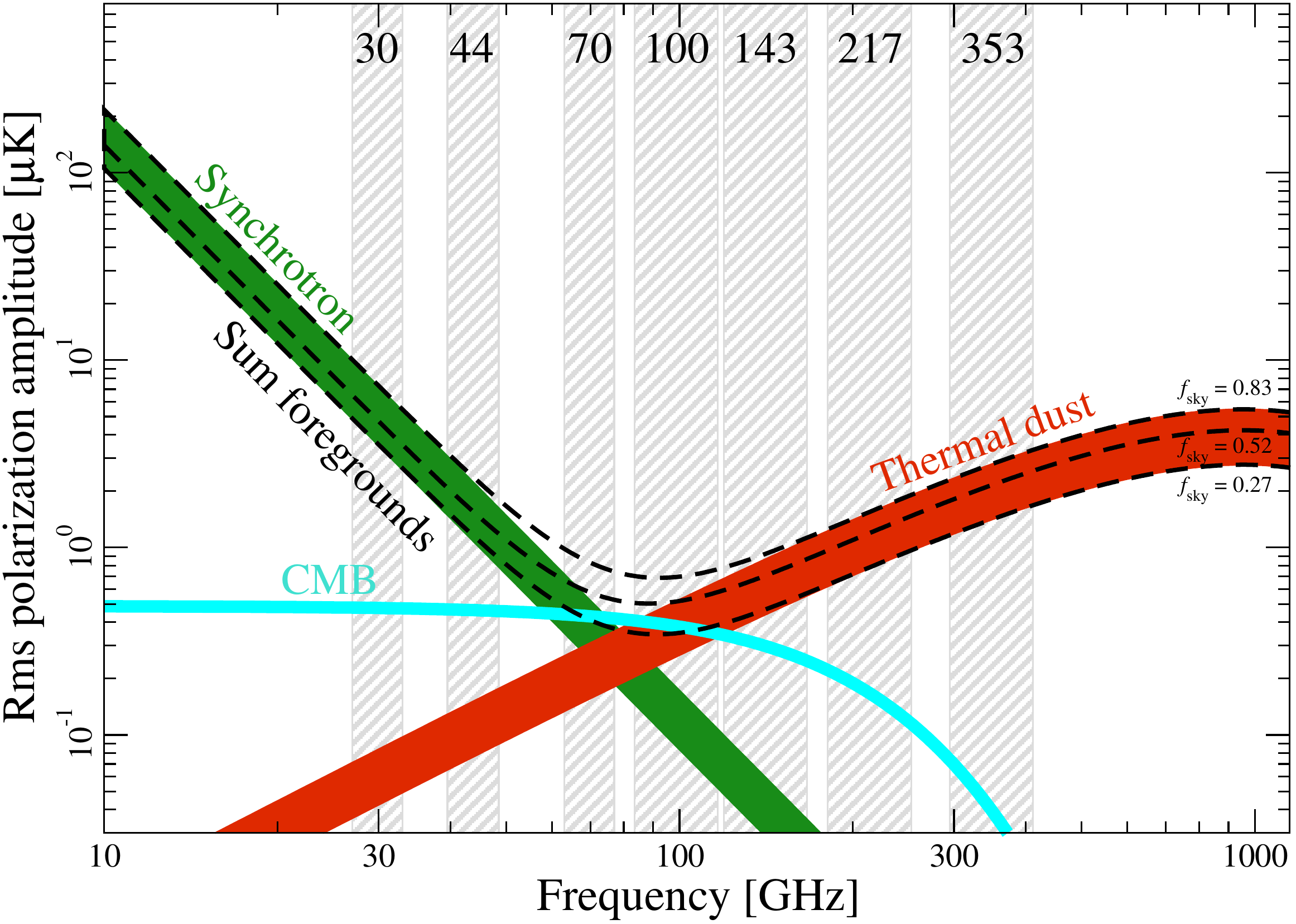}
        \captionof{figure}{CMB and individual foreground contaminants  signal as a function of frequency. Image courtesy of ESA and the Planck Collaboration   \cite{akrami2018planck}.}
        \label{fig:foregrounds_intensity}
    \end{minipage}
    \hspace{.01\linewidth}
    \begin{minipage}[t]{.48\linewidth}
        \centering
        \includegraphics[width=\linewidth]{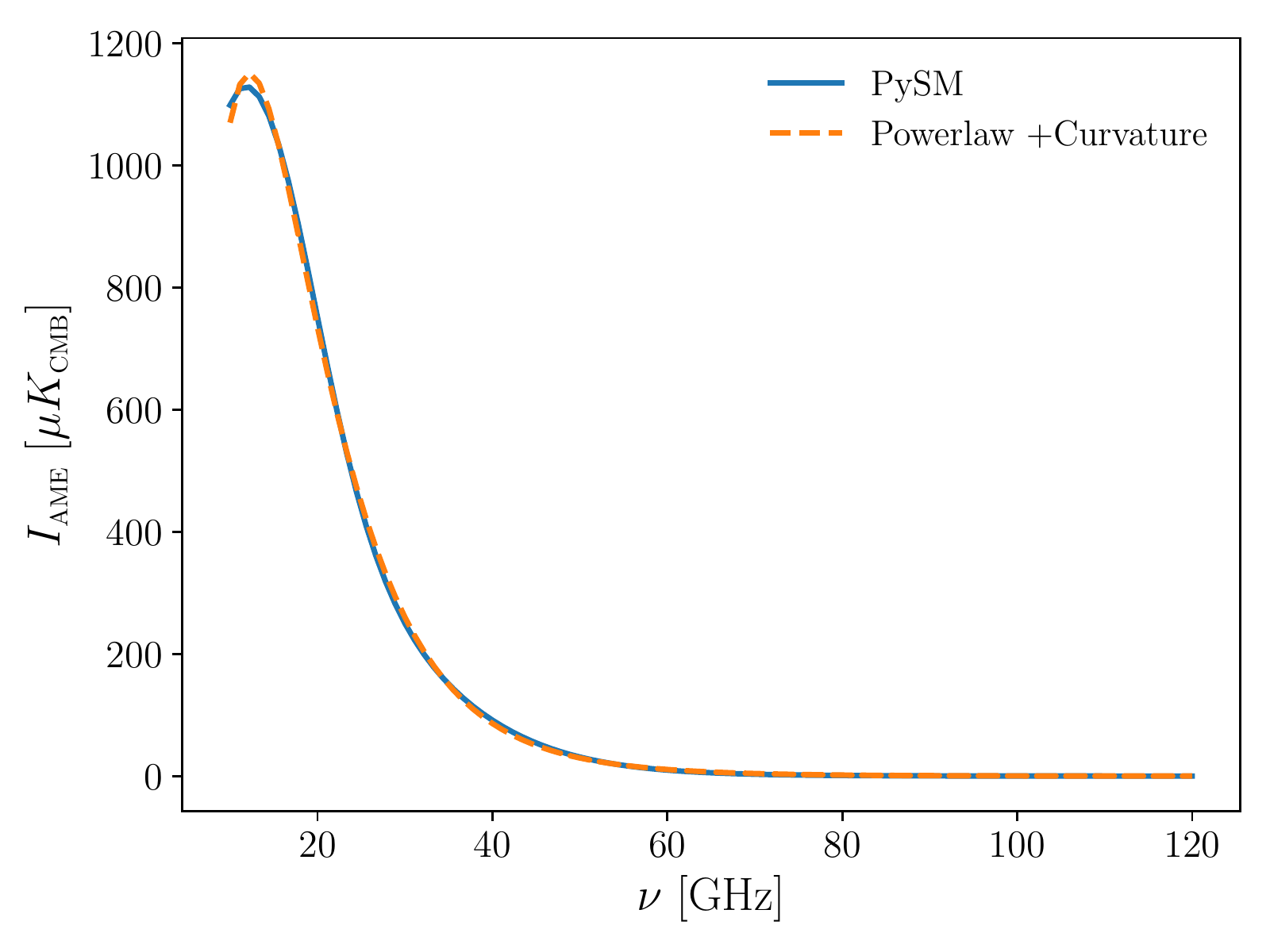}
        \captionof{figure}{Comparison of the spectral behavior of the AME  at a given sky direction, simulated with PySM (solid line), and fitted with a powerlaw with curvature model (dashed line) at the frequencies the LFS operates in.}
        \label{fig:ame_fit}
    \end{minipage}
\end{figure}

\paragraph{Foregrounds:} The polarized foreground contribution is composed primarily of synchrotron and thermal dust\footnote{It is worth mentioning that, at the frequencies we are operating in, Faraday rotation effects are insignificant and can be overlooked.}, as can be seen in figure~\ref{fig:foregrounds_intensity}. However, we have also included the AME in some realizations, as this contaminant might also emit in polarization \cite{genova2015measurements}. The foreground contribution is simulated using parametric models since we want our model and sky simulations to be self-consistent. Below, we describe the procedure followed to create each foreground contribution. Only the Stokes parameters $Q$ and $U$ are considered since we are interested only in polarization.
\begin{itemize}
    \item \textbf{Synchrotron.} This emission is originated from relativistic electrons spiralling around Galactic magnetic fields. Its spectral energy distribution (SED) can be modeled\footnote{The models provided here apply for antenna units. However, we have worked in thermodynamic units using the appropriate change of units where needed.} as a power-law \cite{rybicki2008radiative}. Nevertheless, a model with a curved spectrum might be better suited since it can account for a steepening/flattening of the spectrum due to diverse effects such as: multiple synchrotron components along the line of sight, synchrotron self-absorption, cosmic ray's aging effect, etc. This extension of the model can be thought of as a natural SED's generalization following the approach of \cite{chluba2017rethinking}. Thus, the model used is the following:
    \begin{equation}
        \begin{bmatrix}
		m_{\myvector{n},s}^{\notsotiny{Q}}(\nu;\theta^{\notsotiny{Q}}_{\myvector{n},s})  \\
	    m_{\myvector{n},s}^{\notsotiny{U}}(\nu;\theta^{\notsotiny{U}}_{\myvector{n},s}) 
	\end{bmatrix}
	 = 
	\begin{bmatrix}
	    a_{\myvector{n},s}^{\notsotiny{Q}} \\
	    a_{\myvector{n},s}^{\notsotiny{U}}
	\end{bmatrix}
		\mleft(\dfrac{\nu}{\nu_s}\mright)^{\beta_{\myvector{n},s}+c_{\myvector{n},s}\mleft(\nu/\nu_{s}\mright)} \, ,\\
		\label{eq:synchrotron_model}
    \end{equation}
    where $\myvector{n}$ is a unitary vector pointing in a given direction of the sphere, $m_{\myvector{n},s}^{\notsotiny{X}}$ is the synchrotron signal in the $X$ Stokes parameter ($X \in \{Q,U\}$) at the frequency $\nu$ given $\theta^{\notsotiny{X}}_{\myvector{n},s} =\{a^{\notsotiny{X}}_{\myvector{n},s}, \beta_{\myvector{n},s}, c_{\myvector{n},s}\}$  the set of the synchrotron's model parameters, where $a^{\notsotiny{X}}_{\myvector{n},s}$ is the synchrotron's amplitude at $\nu_s = 23$ GHz, $\beta_{\myvector{n},s}$ is the synchrotron's spectral index, and $c_{\myvector{n},s}$ is the synchrotron's spectral curvature at $\nu_s$. \\
    \\
    The $\map{a}_s^{\notsotiny{X}}$ and $\map{\beta}_s$ template maps at  $n_{side} = 64$ and $256$ were generated  using the template maps of the Python Sky Model (PySM) \cite{thorne2017python}. The maps were degraded from $n_{side} = 512$ to $64$ ($256$) through spherical harmonics, and smoothed with a beam of FWHM = 1$^{\circ}$ ($15$ arcmin), taking into account the pixel window function correction. Besides, latest studies of the galactic synchrotron contribution show that the spectral synchrotron dependence  might have a non-negligible curvature ($c_s = 0.04 \pm 0.1$), \cite{krachmalnicoff2018s}. Thus, we have created a $\map{c}_s$ constant map whose value is $0.04$\footnote{Note that this assumption of a constant value for $\map{c}_s$ does not facilitate its estimation since the method works at the pixel level and spatial correlations are not taken into account.}.
    
    \item \textbf{Thermal Dust.} General name to describe the thermal emission of microscopic matter left in the interstellar space. Dust grains, which are composed mainly of carbonaceous and silicate grains, are heated up by the interstellar radiation field yielding an emission at the microwave range. The dust component SED is well-approximated by a modified black-body \cite{adam2016planck}. However, at the frequencies under study, only the Rayleigh-Jeans part of the dust spectrum is detected, hence a power-law model is also suitable in this particular case:
    \begin{equation}
	\begin{bmatrix}
		m_{\myvector{n},d}^{\notsotiny{Q}}(\nu;\theta^{\notsotiny{Q}}_{\myvector{n},d})  \\
	    m_{\myvector{n},d}^{\notsotiny{U}}(\nu;\theta^{\notsotiny{U}}_{\myvector{n},d})
	\end{bmatrix}
	 = 
	\begin{bmatrix}
		a_{\myvector{n},d}^{\notsotiny{Q}} \\
	    a_{\myvector{n},d}^{\notsotiny{U}}
	\end{bmatrix}
		\mleft(\dfrac{\nu}{\nu_d}\mright)^{\beta_{\myvector{n},d}} \, ,
		\label{eq:dust_model}
	\end{equation}
	where $m_{\myvector{n},d}^{\notsotiny{X}}$ is the dust signal of the Stokes parameter $X$ at the frequency $\nu$ given $\theta^{\notsotiny{X}}_{\myvector{n},d} =\{a^{\notsotiny{X}}_{\myvector{n},d}, \beta_{\myvector{n},d}\}$ the set of the dust's model parameters, where $a^{\notsotiny{X}}_{\myvector{n},d}$ is the dust's amplitude at $\nu_d = 120$ GHz, and $\beta_{\myvector{n},d}$ is the dust's spectral index.\\
	\\
	The $\map{a}_d^{\notsotiny{X}}$ and $\map{\beta}_d$ template maps at  $n_{side} = 64$ and $256$ were created in an analogous manner to the synchrotron's equivalent parameters. 
	
    \item \textbf{AME.} It is a Galactic emission that cannot be explained with known foreground models. Spinning dust grains have been proposed as a mechanism for this emission since it is spatially correlated with dust \cite{stevenson2014derivation,ade2016planck}. Although AME might not be polarized \cite{genova2015measurements}, we have studied some cases were AME contributes to the polarized sky with a 1\% relative amplitude compared to the AME intensity. We have seen that the AME contribution is well-modelled by a power-law with curvature at the frequencies of operation, see figure~\ref{fig:ame_fit}. Therefore the model used is: 
    \begin{equation}
        \begin{bmatrix}
		m_{\myvector{n},a}^{\notsotiny{Q}}(\nu;\theta^{\notsotiny{Q}}_{\myvector{n},a})  \\
	    m_{\myvector{n},a}^{\notsotiny{U}}(\nu;\theta^{\notsotiny{U}}_{\myvector{n},a})
	\end{bmatrix}
	 = 
	\begin{bmatrix}
	    a_{\myvector{n},a}^{\notsotiny{Q}}  \\
	    a_{\myvector{n},a}^{\notsotiny{U}}
	\end{bmatrix}
		\mleft(\dfrac{\nu}{\nu_a}\mright)^{\beta_{\myvector{n},a}+c_{\myvector{n},a}\mleft(\nu/\nu_{a}\mright)} \, ,\\
		\label{eq:ame_model}
    \end{equation}
    where $m_{\myvector{n},a}^{\notsotiny{X}}$ is the AME signal of the Stokes parameter $X$ at the frequency $\nu$ given $\theta^{\notsotiny{X}}_{\myvector{n},a} =\{a^{\notsotiny{X}}_{\myvector{n},a}, \beta_{\myvector{n},a}, c_{\myvector{n},a}\}$ the set of the AME's model parameters, where $a^{\notsotiny{X}}_{\myvector{n},a}$ is the AME's amplitude at $\nu_a = 23$ GHz, $\beta_{\myvector{n},a}$ the AME's spectral index, and $c_{\myvector{n},a}$ is the AME's spectral curvature at $\nu_a$.  \\
    \\
    We obtained maps of the AME's temperature parameters ($\map{a}_a^{I}$, $\map{\beta}_a^{I}$, $\map{c}_a^{I}$) at $n_{side} = 512$ by fitting the PySM default AME's $I$ map to a powerlaw with curvature model. To construct the amplitudes maps in $Q$ and $U$ we have used the dust polarization angles $\map{\gamma}_d$ map, since AME has been shown to be spatially correlated with dust. The amplitudes are then:
    \begin{align}
        \map{a}_{a}^{\notsotiny{Q}} &= \eta \map{a}_{a}^{\notsotiny{I}} \cos(2\map{\gamma}_{d}) \, , &  \map{a}_{a}^{\notsotiny{U}} &= \eta \map{a}_{a}^{\notsotiny{I}} \sin(2\map{\gamma}_{d}) \, , 
    \end{align}
    where $\eta = 0.01$ is the considered AME's ratio of polarization to intensity. $\map{\beta}_a$ and $\map{c}_a$ are the same both in intensity and polarization. Similar to the synchrotron and dust parameters case, the maps were degraded from $n_{side} = 512$ to $64$ ($256$)  through spherical harmonics, and smoothed with a beam of FWHM = 1$^{\circ}$ ($15$ arcmin), taking into account the pixel window function correction. 
\end{itemize}
Note that we have assumed equal spectral parameters for polarization $Q$ and $U$ Stokes parameters.

\paragraph{Noise:} We have included two different types of noise in our simulations: one that consists only of white noise, and another composed of white noise and a correlated noise that resembles the atmospheric and/or systematics contamination.

\begin{itemize}
    \item \textbf{White noise.} The instrument's sensitivity is modeled as a white noise whose standard deviation follows a specific spectral law. The chosen law behaves as the sum of the main foregrounds contaminants in polarization:
    \begin{equation}
    s (\nu) = k_s \mleft(\dfrac{\nu}{100 \, \textrm{GHz}}\mright)^{-3} + k_d \mleft(\dfrac{\nu}{100 \, \textrm{GHz}}\mright)^{1.59} \, ,
    \label{eq:sensitivity}
    \end{equation}
    where we have applied the following constraints to fix $k_s$ and $k_d$:
        \begin{enumerate}
            \item The sensitivity equals $1 \mu $K arcmin at 100 GHz.
            \item The dust-like and synchrotron-like contributions to the sensitivity are equal at 70 GHz.
        \end{enumerate}
    The spectral law is represented in figure~\ref{fig:white_noise_sensitivity}. With this noise behavior, the larger the number of channels the better the effective telescope sensitivity $\bar{s}$, which is defined as:
    \begin{equation}
        \bar{s}_\mathcal{S} = \mleft(\sum\limits_{\nu \in \mathcal{S}}\dfrac{1}{s(\nu)^2}\mright)^{-1/2} \, ,
        \label{eq:effective_sensitivity}
    \end{equation}
    where $\mathcal{S}$ is the set of frequencies in a given setup or band. The default instrument setup [10,10,15] is the largest setup, hence the rest of setups yield always worse results. In order to perform fair comparisons among setups, we have also studied the case where the sensitivity per frequency channel is scaled in the smaller setup to match the default's effective sensitivity. The scaling is conducted by applying the same correction factor $\xi$ to each channels' sensitivity within a band $b$. After applying $\xi$, the smaller setup's effective sensitivity in the $b$ band equals the default's effective sensitivity in the same band, hence
    \begin{equation}
        \xi_b = \sqrt{\dfrac{\bar{s}_{b}^{def}}{\bar{s}_{b}}} \, ,
        \label{eq:corrector_factor_scaling}
    \end{equation}
    where $\bar{s}_{b}^{def}$ and $\bar{s}_{b}$ are the effective sensitivities of the default and smaller setup respectively. 
    \begin{figure}
        \centering
        \begin{minipage}[t]{.48\linewidth}
            \centering
            \includegraphics[width=\linewidth]{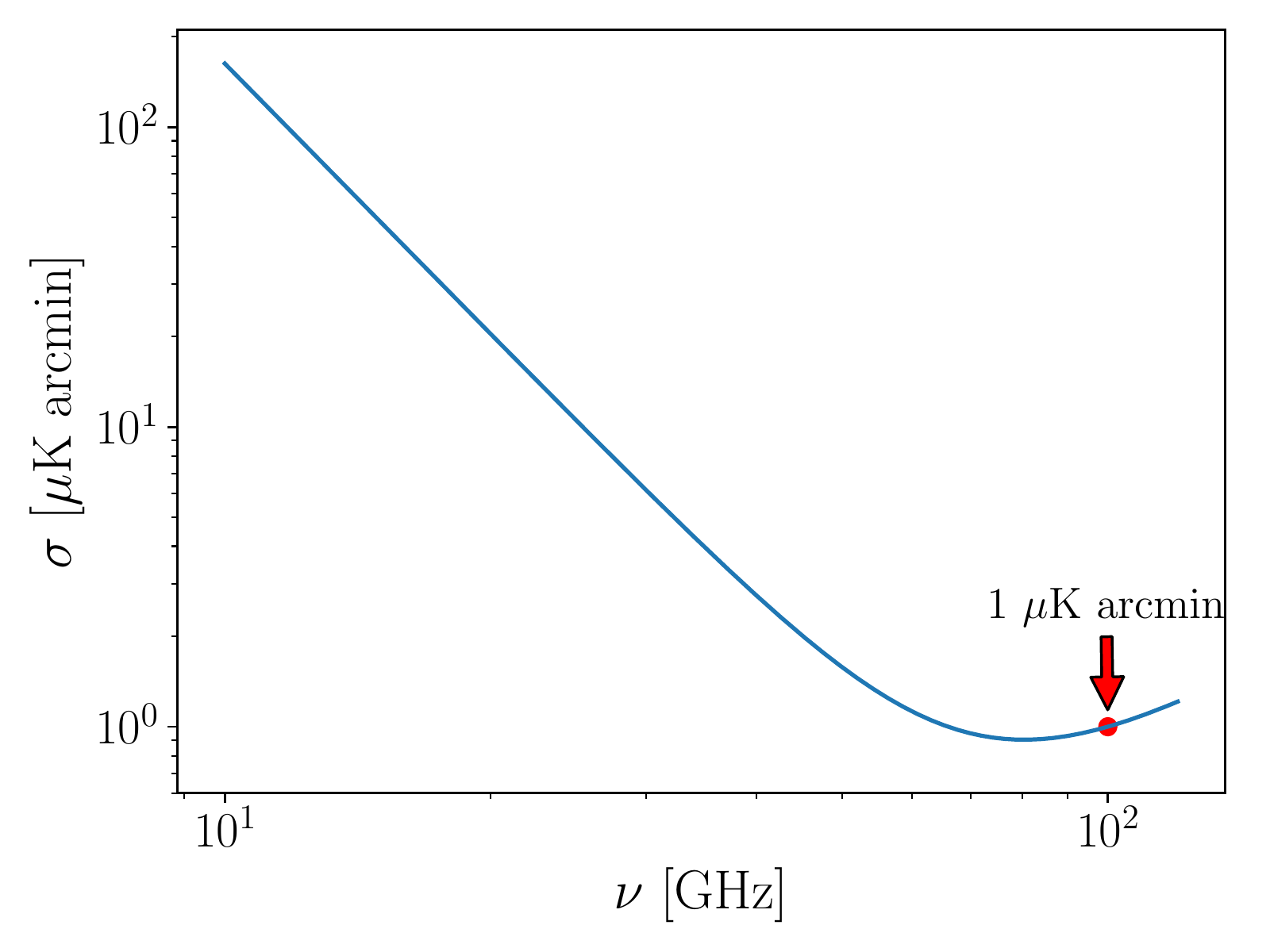}
            \captionof{figure}{Spectral instrument's sensitivity.}
            \label{fig:white_noise_sensitivity}
        \end{minipage}
        \hspace{.01\linewidth}
        \begin{minipage}[t]{.48\linewidth}
            \centering
            \includegraphics[width=\linewidth]{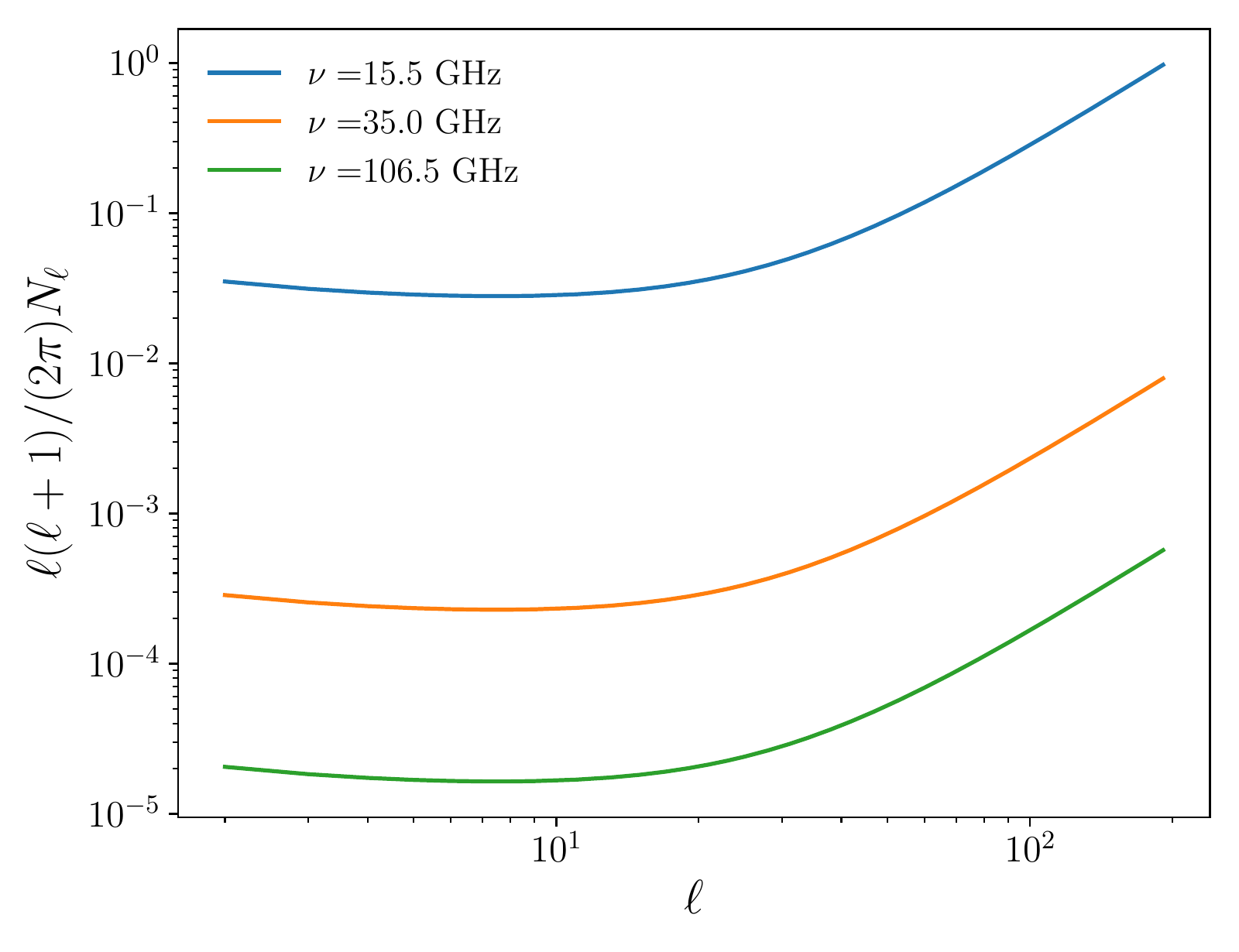}
            \captionof{figure}{Total noise power spectra at different frequencies.}
            \label{fig:white_atmospheric_noise}
        \end{minipage}
    \end{figure}
    \item \textbf{White + Correlated noise.} In this case, a 1/$f$ noise is added to the previously described white noise. This 1/$f$ noise is included to mimic the correlated noise induced by the atmosphere as well as instrument's systematics. This contribution is obtained as a Gaussian random realization of the following power spectrum:
    \begin{equation}
        N_{\ell} = n_{corr}\mleft(\dfrac{\ell}{\ell_{\textrm{knee}}}\mright)^{\gamma}
        \label{eq:atmospheric_power_spectrum}
    \end{equation}
    where $n_{corr}$ is the variance per steradian at a given frequency channel, $\ell_{\textrm{knee}} =30$ is the multipole until which the correlated noise is significantly larger than the white noise contribution, and $\gamma = -2.2, \,-2.4, \, -2.6$ if the frequency channel belongs to the lb, mb or hb respectively. The power spectrum parameters selected are similar to the values considered in \cite{thorne2019removal}. The power spectrum of \eqref{eq:atmospheric_power_spectrum}, along with the white noise contribution, is depicted in figure~\ref{fig:white_atmospheric_noise}.
\end{itemize}

\section{Component Separation}
\label{sec:component_separation}

 Our component separation approach grounds on a full-parametric pixel-based maximum likelihood method, which relies on an affine-invariant ensemble sampler for Markov Chain Monte Carlo (MCMC) \cite{foreman2013emcee}, to retrieve the polarized CMB, as well as the foregrounds' parameters. \\
 \\
 Parametric methods might be more advantageous than non-parametric methods since they provide a physical characterization of both the CMB and the foregrounds. On the other hand, incorrect modeling can lead to severe bias in the measurements in the most extreme cases \cite{armitage2012impact,remazeilles2016sensitivity,fantaye2011estimating}. Nevertheless, there are extensions to these parametric models that can cope with this setback \cite{chluba2017rethinking}. As previously mentioned, our model and sky simulations are self-consistent, i.e., the simulations are generated from the models, hence our results are optimal.\\
 \\
Our method is more robust than other models \cite{fantaye2011estimating,katayama2011simple} since the pixel-based approach employed allows spatial variation of the spectral parameters. The method here is the limiting case considered in \cite{errard2019characterizing} of spatial variability in every single pixel. However, this robustness goes at expense of an increase in the statistical uncertainty of the parameters as less information is provided into the fit \cite{grumitt2019hierarchical}.\\
\\
Hereunder, we outline the application and the fundamentals behind the bayesian inference method employed in this study.

\paragraph{Best-fit Parameter Estimates.}

To obtain sky maps of the most-likely model parameters we used a python implementation \verb|emcee| \cite{foreman2013emcee} of an affine-invariant ensemble sampler for MCMC \cite{goodman2010ensemble}. MCMC methods are algorithms able to sample from a probability distribution, and hence provide an estimation of it. Therefore, we apply this algorithm to draw samples from the global posterior probability and obtain the best-fit parameters' map as the mean of each marginalized parameter posterior probability. The global posterior probability is given by: 
\begin{equation}
    \mathcal{P}(\myset{\theta}\polarization_{\myvector{n}}|\myvector{d}\polarization_{\myvector{n}}) \propto \mathcal{P}(\myvector{d}\polarization_{\myvector{n}}|\myset{\theta}\polarization_{\myvector{n}}) \mathcal{P}(\myset{\theta}\polarization_{\myvector{n}}) \, ,
    \label{eq:posterior}
\end{equation}
 where $\myset{\theta}\polarization_{\myvector{n}}$ is a set whose elements are the $Q$ and $U$ model parameters in a given sky direction $\myvector{n}$, $\myvector{d}\polarization_{\myvector{n}} = (\myvector{d}_{\myvector{n}}^{\notsotiny{Q}},\myvector{d}_{\myvector{n}}^{\notsotiny{U}})$ is a $2n_T$ vector where $\myvector{d}_{\myvector{n}}^{\notsotiny{X}}$ is a $n_T$ vector containing the sky signal $X$ in the experimental setup's $n_T$ frequency channels, $\mathcal{L}(\myset{\theta}\polarization_{\myvector{n}}|\myvector{d}\polarization_{\myvector{n}}) \equiv \mathcal{P}(\myvector{d}\polarization_{\myvector{n}}|\myset{\theta}\polarization_{\myvector{n}})$ is the likelihood function, and $\mathcal{P}(\myset{\theta}\polarization_{\myvector{n}})$ is the prior information known about the parameters. In our approach, $Q$ and $U$ data are jointly fit since they share the spectral model parameters, hence the parameters' statistical uncertainties are reduced. Assuming Gaussian noise, the likelihood of the data can be expressed as 
\begin{equation}
    \mathcal{L}(\myset{\theta}\polarization_{\myvector{n}}|\myvector{d}\polarization_{\myvector{n}}) = \dfrac{1}{\sqrt{(2\pi)^{2N}\det (\mymatrix{C})}}\exp\mleft(-\dfrac{1}{2}\mleft(\myvector{d}\polarization_{\myvector{n}}-\mathbf{m}\polarization\mleft(\myvector{\nu};\myset{\theta}\polarization_{\myvector{n}}\mright)\mright)^{T}\mymatrix{C}^{-1}\mleft(\myvector{d}\polarization_{\myvector{n}}-\mathbf{m}\polarization\mleft(\myvector{\nu};\myset{\theta}\polarization_{\myvector{n}}\mright)\mright)\mright) \, ,
    \label{eq:likelihood}
\end{equation}
where $\mymatrix{C} = diag(\mathbf{C}^{\textrm{\notsotiny{Q}}},\mymatrix{C}^{\notsotiny{U}})$  being $\mymatrix{C}^{\notsotiny{X}}$ the covariance matrix of the the telescope's frequency channels for the Stokes parameter $X$\footnote{In this work we have assumed $\mymatrix{C}^{\notsotiny{Q}} = \mymatrix{C}^{\notsotiny{U}}$ }, and $\myvector{m}\polarization = (\mathbf{m}^{\notsotiny{Q}},\myvector{m}^{\notsotiny{U}})$
a $2n_T$ vector containing the model signal, and:
\begin{equation}
    \mathbf{m}^{\notsotiny{X}}\mleft(\nu;\myset{\theta}^{\notsotiny{X}}_{\myvector{n}}\mright) = c_\myvector{n}^{\notsotiny{X}} + \sum\limits_{f \in F}\myvector{m}_{\myvector{n},f}^{\notsotiny{X}}\mleft(\nu;\myset{\theta}^{\notsotiny{X}}_{\myvector{n},f}\mright) \, ,
    \label{eq:parametric_model}
\end{equation}
where $c_\myvector{n}^{\notsotiny{X}}$ is the CMB $X$-contribution in the direction $\myvector{n}$, $F$ is the set of foregrounds included in a given model, e.g, $F = \{s,d\}$ in a model with only synchrotron and thermal dust, and  $\myvector{m}_{\myvector{n},f}^{\notsotiny{X}}(\myset{\theta}^{\notsotiny{X}}_{\myvector{n},f})$ is a vector whose elements are the $f$ foreground model contribution at a given frequency obtained by evaluating the $f$ foreground parametric model \eqref{eq:synchrotron_model}-\eqref{eq:ame_model}, using the set of model parameters $\myset{\theta}^{\notsotiny{X}}_{\myvector{n},f}$.\\
\\
Priors are required in Bayesian inference and have been proven to help with convergence and computational time reduction. In this analysis we have used Gaussian priors:
    \paragraph{Gaussian priors.} We have applied Gaussian priors to the spectral parameters. Gaussian priors are given by:
    \begin{equation}
        \mathcal{P} (\theta_{\myvector{n}}) = \exp\mleft(-\dfrac{1}{2}\dfrac{(\theta_{\myvector{n}}-\mu_{\theta})^2}{\sigma_{\theta}^2}\mright)
        \label{eq:gaussian_prior}
    \end{equation}
    where $\theta$ is a given model parameter, and,  $\mu_{\theta}$ and $\sigma_{\theta}^2$ are the mean and variance of the parameter $\theta$. The means and standard deviations used are listed in table~\ref{tab:priors}. Notice that we have used the spectral parameters template maps 3-$\sigma$ values as $\sigma_{\theta}$ to loosen the priors. 

\begin{table}
    \centering
    \begin{tabular}{|c|ccccc|}
    \hline
        &   $\beta_s$ &   $\beta_d$ & $\beta_a$ & $c_s$ & $c_a$ \\
    \hline
        $\mu_{\theta}$ & -3.00  & 1.54 & -2.5 & 0.04 & -2.0\\
        $3\sigma_{\theta}$ & 0.18 & 0.12 & 2.1 & 0.10 & 0.9\\
    \hline
    \end{tabular}
    \caption{\textbf{Gaussian prior information.} Displayed are the mean and the dispersion values of the spectral parameters employed in the Gaussian priors. $\mu_{\theta}$ and $\sigma_{\theta}$ are the mean and the 3-$\sigma$ value of $\theta$ template map respectively.}
    \label{tab:priors}
\end{table}

\section{Residual Power Spectra Estimation}
\label{sec:self-residual_estimation}

In this study, we have developed a self-consistent approach to obtain an estimate of the combined foreground and instrumental model residuals power spectrum. Having a residuals model allow us to prevent possible biases in the fit due to an insufficient foreground removal, or to determine the range of detectable $r$ values, given a specific experimental setup. Another advantage of this methodology is that it can be applied to real data. In this section, we describe the methodology followed to calculate the residuals model (section~\ref{subsec:residuals_model_estimation}), explain the approach used to estimate the cosmological parameters (section~\ref{subsec:cosmological_parameters}), and show an example for the default scenario (section~\ref{subsec:example_default_scenario}). A scenario is fixed when the following characteristics are set: i. the experimental setup, i.e., the 3-tuple [$n_{lb}$,$n_{mb}$,$n_{hb}$]; ii. the cosmological parameters that define the $B$-mode power spectrum ($r$, $a_L$); iii. the foreground model, i.e., the specific foregrounds that are included; iv. the noise type; and v. the maps' resolution, i.e., $n_{side}$. The scenarios characteristics are listed in table~\ref{tab:scenarios}.
\begin{table}
    \centering
    \begin{tabular}{|cccccccc|}
    \hline
       Scenario &   setup &   $r\times 10^{3}$ & $a_L$ & $F$ & noise & $n_{side}$  & sky \\
    \hline
        default & [10,10,15]  & 0 & 1 & {s,d} & W & 64 & P70\\
        558 & [5,5,8]  & 0 & 1 & {s,d} & W & 64 & P70\\
        558-scaled & [5,5,8]  & 0 & 1 & {s,d} & WS & 64 & P70\\
        666 & [6,6,6]  & 0 & 1 & {s,d} & W & 64 & P70\\
        666-scaled & [6,6,6]  & 0 & 1 & {s,d} & WS & 64 & P70\\
        default-delensed & [10,10,15]  & 0 & 0.5 & {s,d} & W & 64 & P70\\
        starobinsky & [10,10,15]  & 3.7 & 1 & {s,d} & W & 64 & P70\\
        starobinsky-delensed & [10,10,15]  & 3.7 & 0.5 & {s,d} & W & 64 & P70\\
        AME & [10,10,15]  & 0 & 1 & {s,d,a} & W & 64 & P70\\
        NH & [10,10,15]  & 0 & 1 & {s,d} & W & 64 & Q(WS) \\
        cosmoareas & [10,10,15]  & 0 & 1 & {s,d} & W($f_{sky}$) & 256 & Q(CA)\\
        correlated-noise & [10,10,15]  & 0 & 1 & {s,d} & W+Corr & 64 & P70\\
        LB & LB  & 0 & 1 & {s,d} & W & 64 & P70\\
        LB/LFS & LB+[10,10,0]  & 0 & 1 & {s,d} & W & 64 & P70\\
        LB-AME & LB  & 0 & 1 & {s,d,a} & W & 64 & P70\\
        LB/LFS-AME & LB+[10,10,0]  & 0 & 1 & {s,d,a} & W & 64& P70\\
    \hline
    \end{tabular}
    \caption{\textbf{Scenarios.} The different studied scenarios' characteristics. The setup 3-tuple are the number of channels in each frequency band in the LFS, while LB stands for LiteBIRD's frequency channels; $r$ and $a_L$ are the input tensor-to-scalar ratio and lensing amplitude; $F$ is the set of foregrounds included in the sky signal simulation, where $s$, $d$, and $a$ stands for synchrotron, thermal dust, and AME respectively; the noise included in the simulations are white noise (W), scaled white noise (WS), white plus correlated noise (W+Corr) and, white noise with a scaling factor to account for the longer observational time spent when only small patches of the sky are measured (W($f_{sky}$)); $n_{side}$ is the resolution of the simulated signal maps; and finally sky specifies the observable sky studied in each scenario, P70, Q(WS) and Q(CA) are the sky left after applying the \texttt{Planck}, \texttt{QUIJOTE} wide survey combined with \texttt{Planck}, and \texttt{QUIJOTE} cosmological areas mask respectively.}
    \label{tab:scenarios}
\end{table}

\subsection{Residuals Model Estimation}
\label{subsec:residuals_model_estimation}

Let the data polarization signal be defined as $\mymatrix{S}\polarization= (\mymatrix{S}^{\notsotiny{Q}},\mymatrix{S}^{\notsotiny{U}})$,  where $\mymatrix{S}^{\notsotiny{X}} = (\map{s}^{\notsotiny{X}}_{\nu_1} \cdots \, \map{s}^{\notsotiny{X}}_{\nu_{n_T}})$ is a ($n_{pix} \times n_{T}$) matrix whose columns are the $X$-signal maps $\map{s}_{\nu_{j}}^{\notsotiny{X}}$ at the frequency $\nu_{j}$, being $n_{pix}$ the number of map pixels. Let $\myvector{d}\polarization_{p}$ be a $\mymatrix{S}\polarization$ row, i.e., $\myvector{d}\polarization_{\myvector{n}}$ with $\myvector{n}$ pointing in the direction of the pixel $p$. The procedure to obtain the residuals model estimate is the following:
\begin{enumerate}
    \item For each pixel $p$, the best-fit set of model parameters $\myset{\theta}^{dat}_{p}$ is obtained by applying the Bayesian method explained in section~\ref{sec:component_separation} to $\myvector{d}_p$. Eventually, we obtain $\maparray{\Theta}^{dat}$ a ($n_{pix}\times n_{par}$) matrix whose rows are the $\myset{\theta}^{dat}_{p}$ and, its columns are the model parameters' maps. $n_{par}$ is the number of model parameters.  
    \item Then, $n_{sim}$ signal matrices $\{\mymatrix{S}\polarization_{j}\}_{j \in \{1,...,n_{sim}\}}$ are generated using 
    \begin{equation}
        \myvector{d}^{\notsotiny{X}}_{p,j}(\nu) = c^{\notsotiny{X}}_{p,j} + \sum\limits_{f \in F}\myvector{m}_{p,f}^{\notsotiny{X}}\mleft(\nu;\myset{\theta}^{\notsotiny{X},dat}_{p,f}\mright) + \myvector{n}^{\notsotiny{X}}_{p,j}(\nu) \, ,
        \label{eq:data_simulation}
    \end{equation}
    where $c_{p,j}^{\notsotiny{X}}$ is the pixel $p$ value of the $j$-th simulated CMB map $\map{c}_{j}^{X}$, generated as a Gaussian random realization of a particular set of power spectra, the second term of the right-hand-side of \eqref{eq:data_simulation} is a vector containing the foregrounds contribution obtained by evaluating the $f$ foreground parametric model using the estimated $\theta^{\notsotiny{X},dat}_{p,f}$ as model parameters, and $\myvector{n}^{\notsotiny{X}}_{p,j}$ is a noise vector obtained as a random realization of the noise model.
    \item Step 1. is repeated for each $\mathbf{S}\polarization_{j}$ to retrieve $\maparray{\Theta}^{sim}_{j}$. 
    \item Next, the CMB $X$ Stokes parameter residual maps are calculated for each $j$ simulation as 
    \begin{equation}
        \map{c}^{X,res}_{j} = \map{c}^{X}_{j} - \map{c}^{X,sim}_{j} \, ,
        \label{eq:cmb_residual_map}
    \end{equation}
    while the foreground residuals maps at a given frequency $\nu$ is given by
    \begin{equation}
        \map{m}^{X,res}_{f,j} = \map{m}^{X}_{f,j}\mleft(\nu;\map{\theta}^{X,dat}_{f}\mright) - \map{m}^{X}_{f,j}\mleft(\nu;\map{\theta}^{X,sim}_{f,j}\mright) \, .
        \label{eq:foreground_residual_map}
    \end{equation}
    \item Finally, for each $j$ residual map, the power spectra is obtained using a  pseudo-$C_{\ell}$ algorithm \cite{wandelt2001cosmic,hivon2001master}. Pseudo-$C_{\ell}$ algorithms are a technique to solve the $E$-to-$B$ leakage  due to the scale spherical harmonics mixing in partial-sky maps\footnote{Note that in this work, only partial sky maps are studied since we always apply a galactic mask.}. Even though this method does not retrieve the minimum variance \cite{tegmark1997measure}, it is the most broadly used approach since it is not computationally expensive. Moreover, there are techniques to reduce the $B$ variance due to the overpowering $E$-to-$B$ mode leakage, like the ``pure'' pseudo-$C_{\ell}$ mechanism \cite{bunn2003b}. This mechanism requires the mask to satisfy both the Neumann and Dirichlet conditions \cite{alonso2019unified}. The latter is achieved by apodizing the mask, i.e., artificially making the mask's edges less abrupt. In this work we have employed a ``pure'' pseudo-$C_{\ell}$ algorithm using the python implementation of the public software package \verb|NaMaster| \cite{alonso2019unified}. The residual model power spectrum estimate $R_{\ell}$ is calculated as the mean of the $n_{sim}$ residuals power spectra.
\end{enumerate}

\subsection{Cosmological Parameters Fit}
\label{subsec:cosmological_parameters}

The cosmological parameters can be estimated by fitting the power spectrum of the cleaned map $\map{c}^{dat}$ to the theoretical primordial and lensing CMB power spectra as well as the residuals model power spectrum. In this work we are interested mainly in $r$, hence we only conduct the analysis on the $BB$ power spectrum. Since only partial-sky maps are studied, the large scale multiples cannot be accessed. Thus, in the limit of high enough multipoles, the likelihood of the cosmological parameters can be approximated to a Gaussian:
\begin{equation}
    -\log \mathcal{L}(r,a_L,a_R) \propto \sum\limits_{\ell} \dfrac{\mleft(C_{\ell}^{dat} - r B^{GW}_{\ell}(r=1) - a_L L_{\ell} - a_R  R_{\ell} \mright)^{2}}{\sigma_{\ell}^2} \, ,
    \label{eq:loglikelihood_r}
\end{equation}
where $C_{\ell}^{dat}$ is the $B$-mode power spectrum of the best-fit CMB map, $B_{\ell}^{GW}$ is the $B$-mode  primordial power spectrum at $r=1$, $L_{\ell}$ is the lensing contribution to the $BB$ power spectrum, $R_{\ell}$ is the residuals model $BB$ power spectrum, and $\sigma_{\ell}$ is the cosmic variance:
\begin{equation}
    \sigma_{\ell} = \sqrt{\dfrac{C_{\ell}^2}{f_{sky}\mleft(\ell + \frac{1}{2}\mright)}} = \dfrac{rB^{GW}_{\ell}(r=1)+a_L L_{\ell} + a_R R_{\ell}}{f_{sky}^{1/2}\mleft(\ell + \frac{1}{2}\mright)^{1/2}} \, .
\end{equation}
In order to maintain the same statistical properties as well as the binning (required to perform the Pseudo-$C_{\ell}$ algorithm), we have generated  $B^{GW}_{\ell}$ and  $L_{\ell}$ models as the mean of the mask corrected power spectra of 100 realizations of the theoretical $B^{GW}_{\ell}$ and  $L_{\ell}$ respectively.\\
\\
By minimizing \eqref{eq:loglikelihood_r}  the best-fit $r$, $a_L$ and, $a_R$ parameters can be derived analytically, and their uncertainties can be evaluated from the Fisher matrix. 
It is worth noting that with this approach we marginalize over $a_R$ and $a_L$ which leads to more conservative results than fixing those parameters to unity.

\subsection{Example: Default Scenario}
\label{subsec:example_default_scenario}

In this section, the default scenario's results are shown to: i. validate the method's self-consistency, i.e., assess whether the residuals model obtained reproduces the true residuals, and ii. present some of the method's results. In this analysis we have applied only a galactic disk mask, i.e.,  the \verb|Planck|'s mask fig.~\ref{subfig:planck_mask}. 

\paragraph{Self-consistency:} To test the validity of our method we have compared the power spectra of the true residuals with the residuals obtained by simulations.  
\begin{figure}
    \centering
    \includegraphics[width=\linewidth]{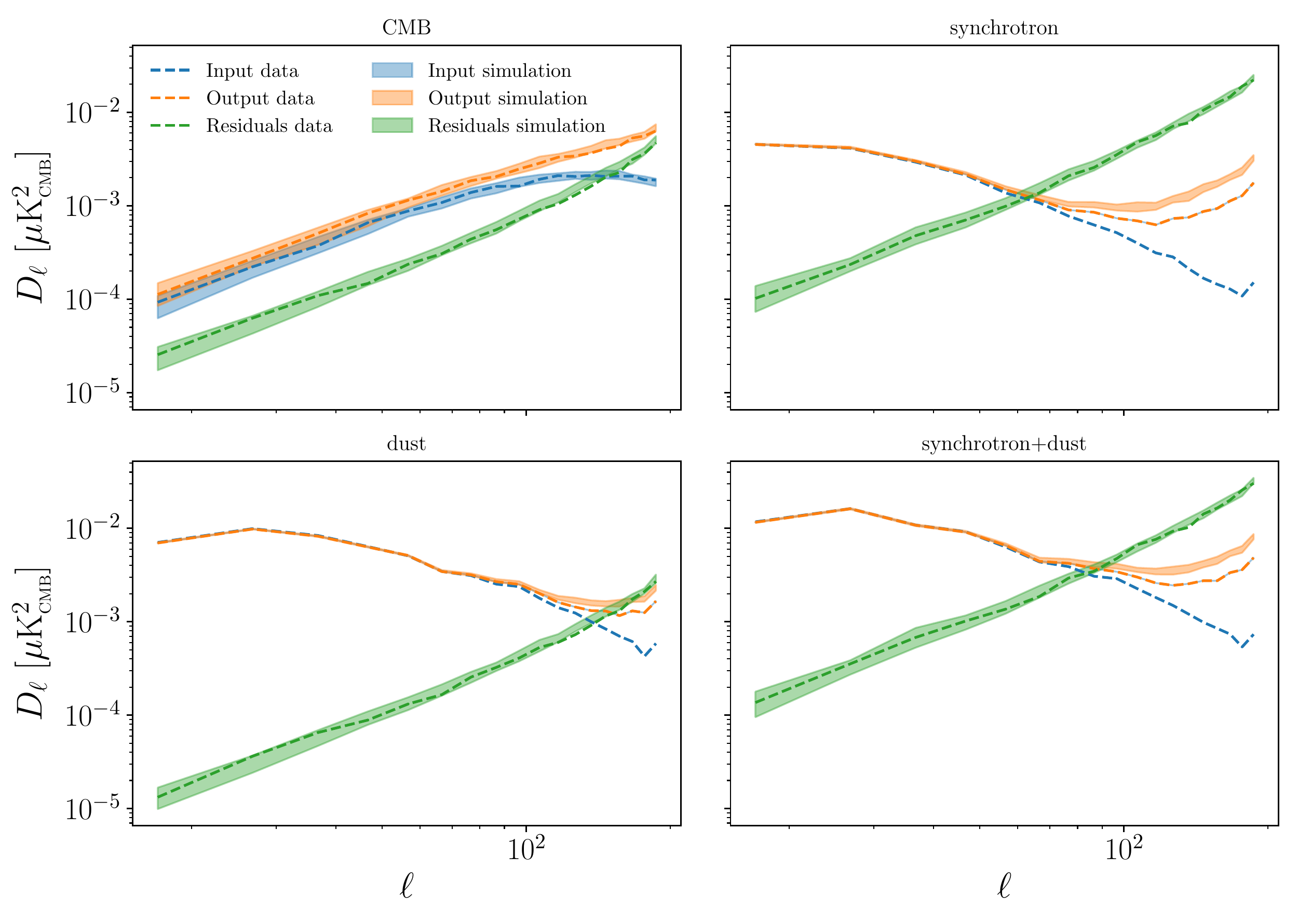}
    \caption{Power spectrum of the input, output, and residuals best-fit data versus simulation of different sky components. The simulation power spectra are represented as colored shaded areas, while the data are depicted with colored dashed lines. The foregrounds power spectra are evaluated at $\nu = 76.5$ GHz.} 
    \label{fig:simulation_vs_data_power_spectra}
\end{figure}
This is performed in figure~\ref{fig:simulation_vs_data_power_spectra} where the input, output, and residuals power spectra of both the data and the simulations are depicted together. Let us analyze the CMB and the foregrounds separately. 
\begin{itemize}
    \item \textbf{CMB.} We observe in figure~\ref{fig:simulation_vs_data_power_spectra} that the data and simulations input are very similar which is expected as they are generated  in a similar fashion. In the case of real data, the true cosmological values are not known, but one can use $r=0$ since the residuals left are substantially larger than the primordial signal. However, in this particular case we have used the same $r$ for both the simulations and the data. Furthermore, we observe that the simulations output, and consequently, the residuals agree with the data output and residuals. Thus, we are confident that with this method coherent results can be obtained.
    \item \textbf{Foregrounds.} The input simulated foreground parameters are the best-fit parameters of the data. This is the reason why the input simulated power spectra overlap with the data output power spectrum. On the other hand, we observe in figure~\ref{fig:simulation_vs_data_power_spectra} that there is a slight increase in the simulated output power spectra at small scales. The data estimated foreground parameters already have an uncertainty since they are retrieved from noisy data. These foregrounds parameters are used to create new noisy sky simulations, hence the simulated output foreground parameters will have a larger uncertainty compared to the data foreground parameters. Regardless of this difference, the simulation residuals resembles the true residuals. Therefore, the mean simulation residuals is a suitable approximation of the true residuals. 
\end{itemize}

\paragraph{Results:} 

\begin{figure}
    \centering
    \includegraphics[width=.6\linewidth]{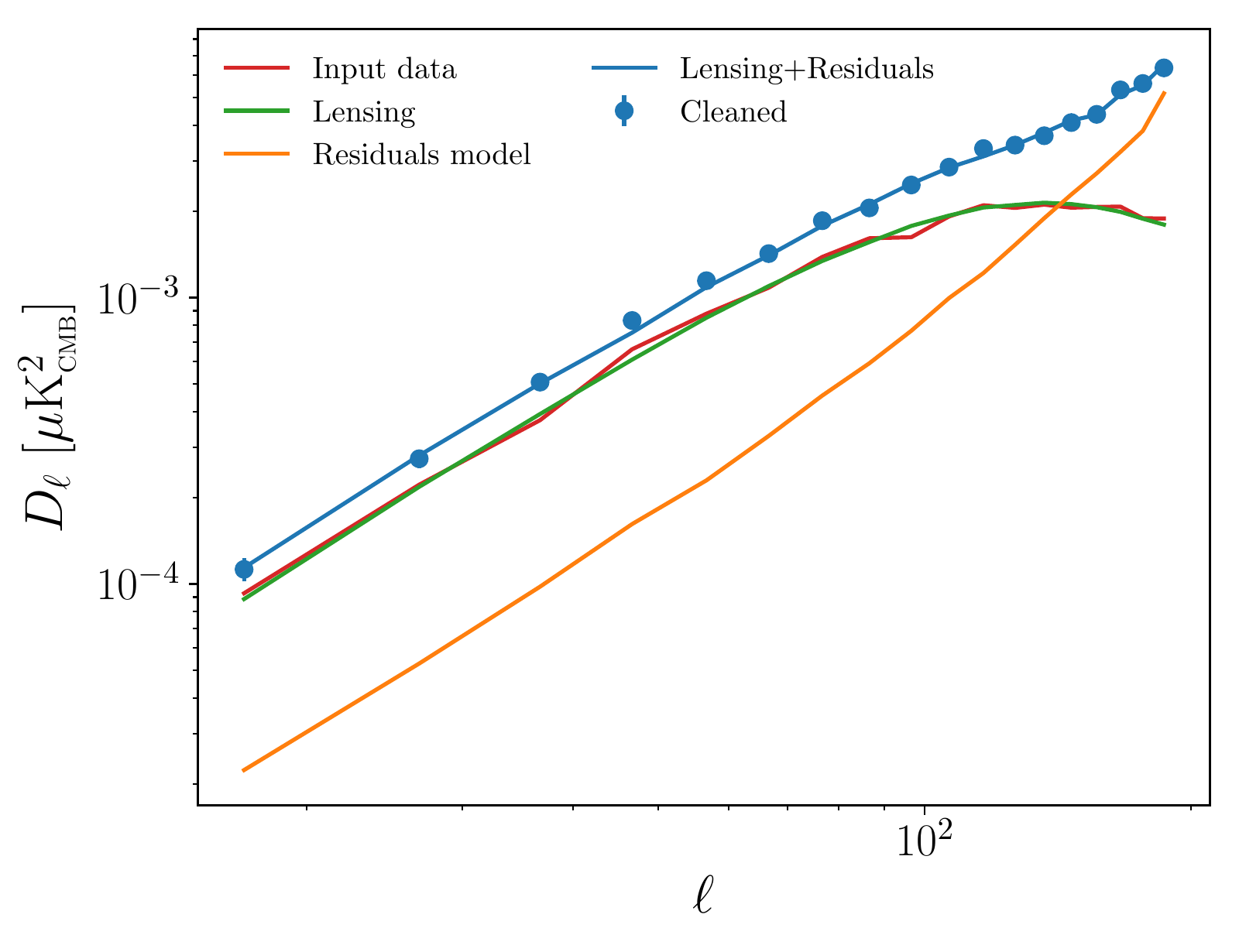}
    \caption{CMB $B$-mode power spectra contributions in the default case. Note that the primordial $B$-mode is not shown since $r=0$ in the default scenario.}
    \label{fig:plot_results_default}
\end{figure}
\begin{figure}
    \centering
    \includegraphics[width=.6\linewidth]{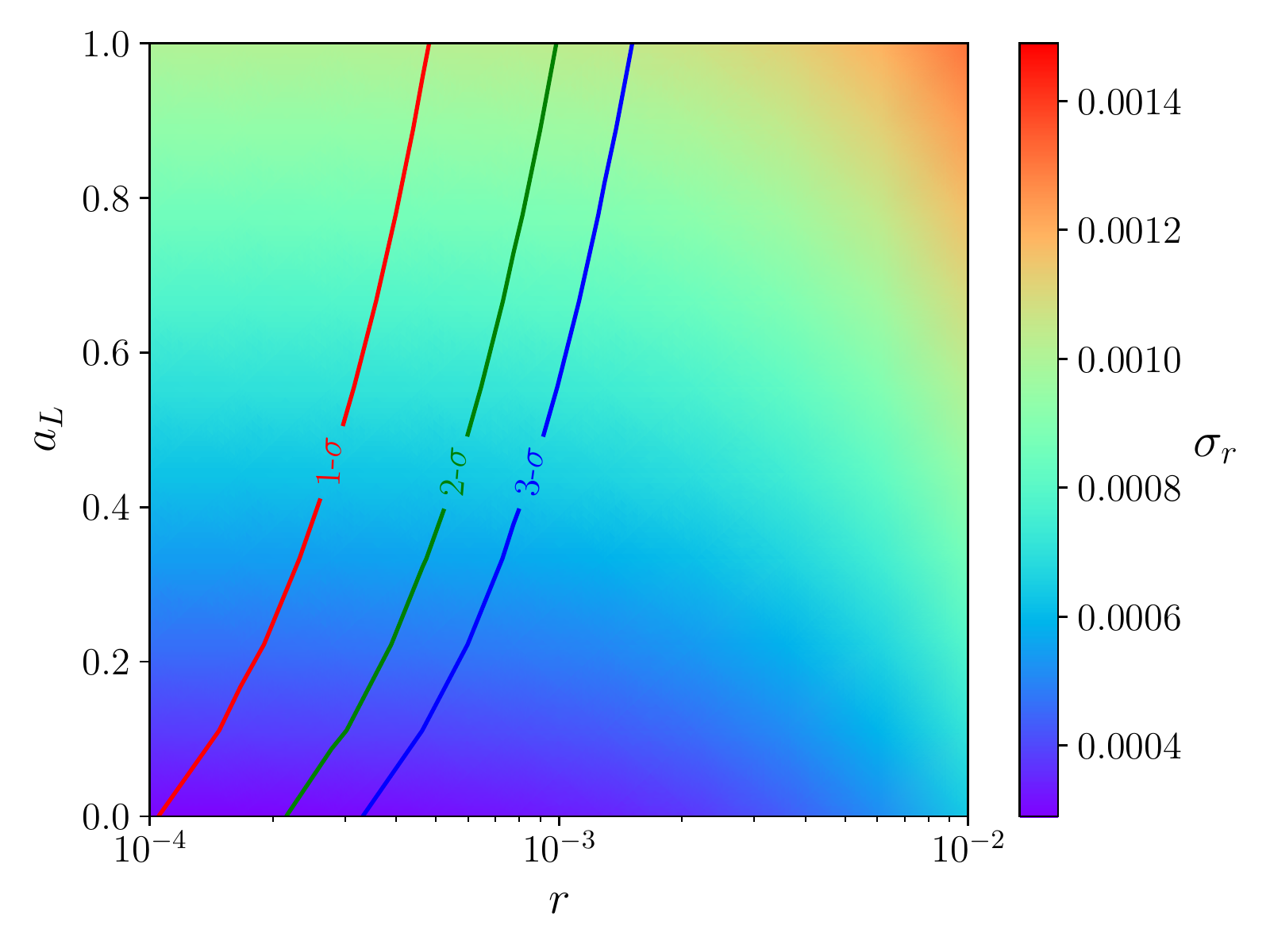}
    \caption{Density plot of $\sigma_r$ as a function of $r$ and $a_L$, for fixed $a_R=1$ with the default's residuals model. The 1,2,3-$\sigma$ signal-to-noise contours are also represented.}
    \label{fig:sigma_r_density_plot_default}
\end{figure}

Once the default scenario residuals model $R_{\ell}$ is obtained we fit the $BB$ power spectrum of the cleaned CMB map as described in section~\ref{subsec:cosmological_parameters}. In figure~\ref{fig:plot_results_default} we show the $BB$ power spectra of the different components after correcting for the mask leakage. It is clear from figure~\ref{fig:plot_results_default} that the CMB lensing is the main source of uncertainty except at small scales, where the foreground residuals become dominant. Thus, if $r$ is sufficiently small delensing would be mandatory in order to make a detection. 
\begin{figure}
    \centering
    \includegraphics[width=\linewidth]{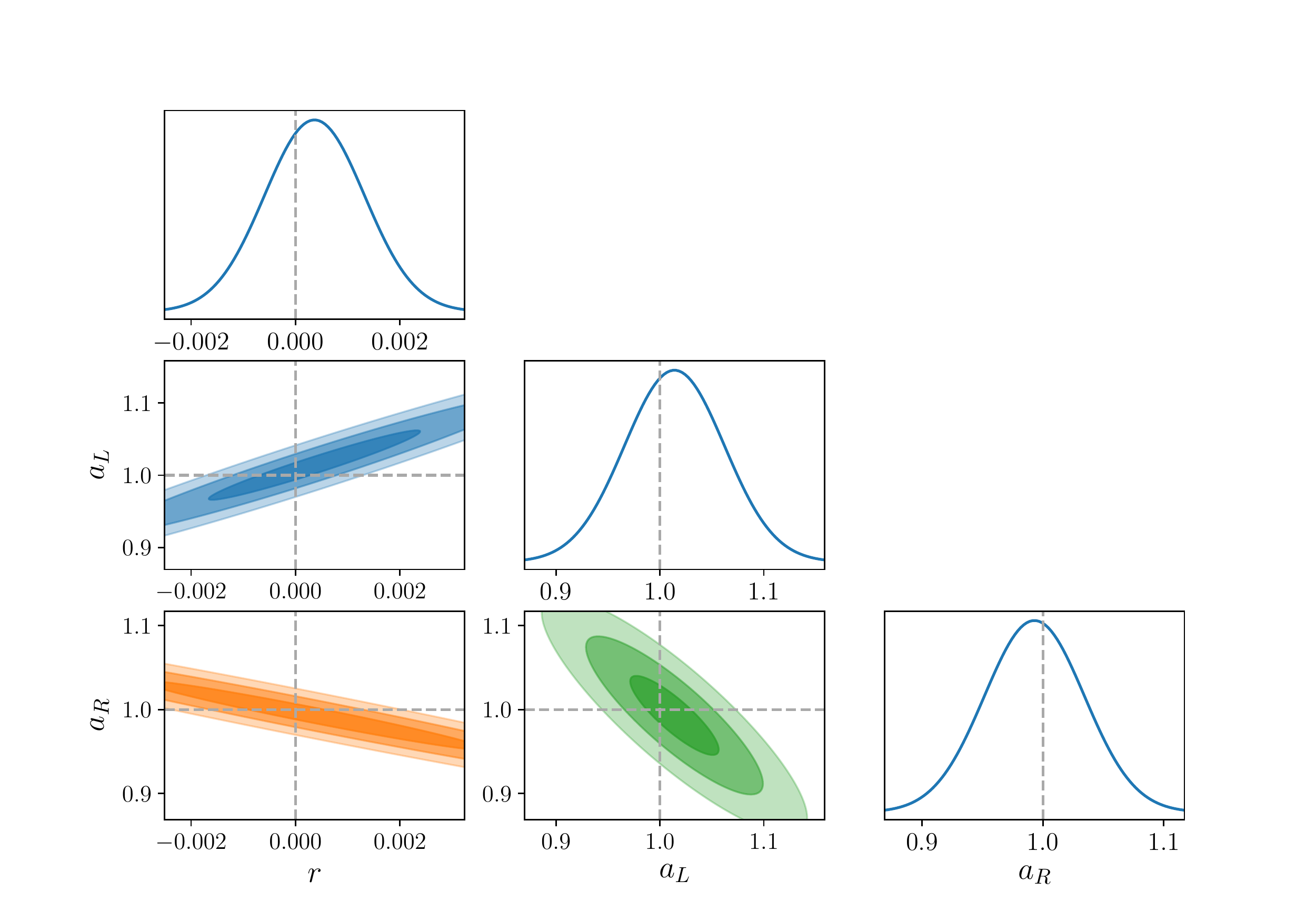}
    \caption{Marginalized Gaussian one and two dimensional projections of the posterior distribution of the fit parameters.}
    \label{fig:triangular_plot_default}
\end{figure}\\
\\
The Gaussian distributions and covariances among the parameters obtained from the fit are shown in figure~\ref{fig:triangular_plot_default}, and the numerical results in table~\ref{tab:scenario_results}. $r$, $a_L$ and $a_R$ values are compatible with their true value and  $\sigma_r \lesssim 10^{-3}$ which is the target value of most experiments. From the covariance matrices it is clear that each parameter is correlated with the rest. Therefore a bias on the $r$ value could yield a mismatch between $a_L$ and $a_R$ from unity. If the residuals and lensing model are appropriate, $a_L$ and $a_R$ departures from the true value can be used  to detect biases on the recovered $r$.
\\
\\
Moreover, given the residuals model an estimation of the $r$ uncertainty can be obtained using the following equation:
\begin{equation}
    \sigma_r = \mleft(\sum\limits_{\ell}\dfrac{{B_{\ell}^{GW}}^2(r=1)}{\sigma_{\ell}^2}\mright)^{-1/2} \, ,
    \label{eq:sigma_r}
\end{equation}
which is derived as the marginal uncertainty of the parameter $r$ of the covariance matrix, calculated from the Fisher matrix. $\sigma_r$ depends on the value of $r$, $a_L$ and $a_R$ through $\sigma_{\ell}$. In figure~\ref{fig:sigma_r_density_plot_default} $\sigma_r$ is depicted as a function of $r$ and $a_L$ having $a_R$ fixed to unity.
\\
\\
From figure~\ref{fig:sigma_r_density_plot_default} we infer that within this scenario values of $r\lesssim 10^{-3}$ are not detectable, i.e., signal to noise ratio larger than 3-$\sigma$, without performing any form of delensing.

\section{Experiment Performance}
\label{sec:experiment_performance}

In this section the LFS performance is analyzed in a handful of different scenarios where the following characteristics are tested: the experimental setup (section~\ref{subsec:instrument_setup}); different cosmologies (section~\ref{subsec:primordial_b_modes}); the inclusion of AME  (section~\ref{subsec:foreground_model}); different observation strategies (section~\ref{subsec:observation_strategies}); and the atmospheric/systematics contamination (section~\ref{subsec:atmospheric_contamination}). 

\subsection{Instrumental Setup}
\label{subsec:instrument_setup}

\begin{figure}
    \centering
    \includegraphics[width=.6\linewidth]{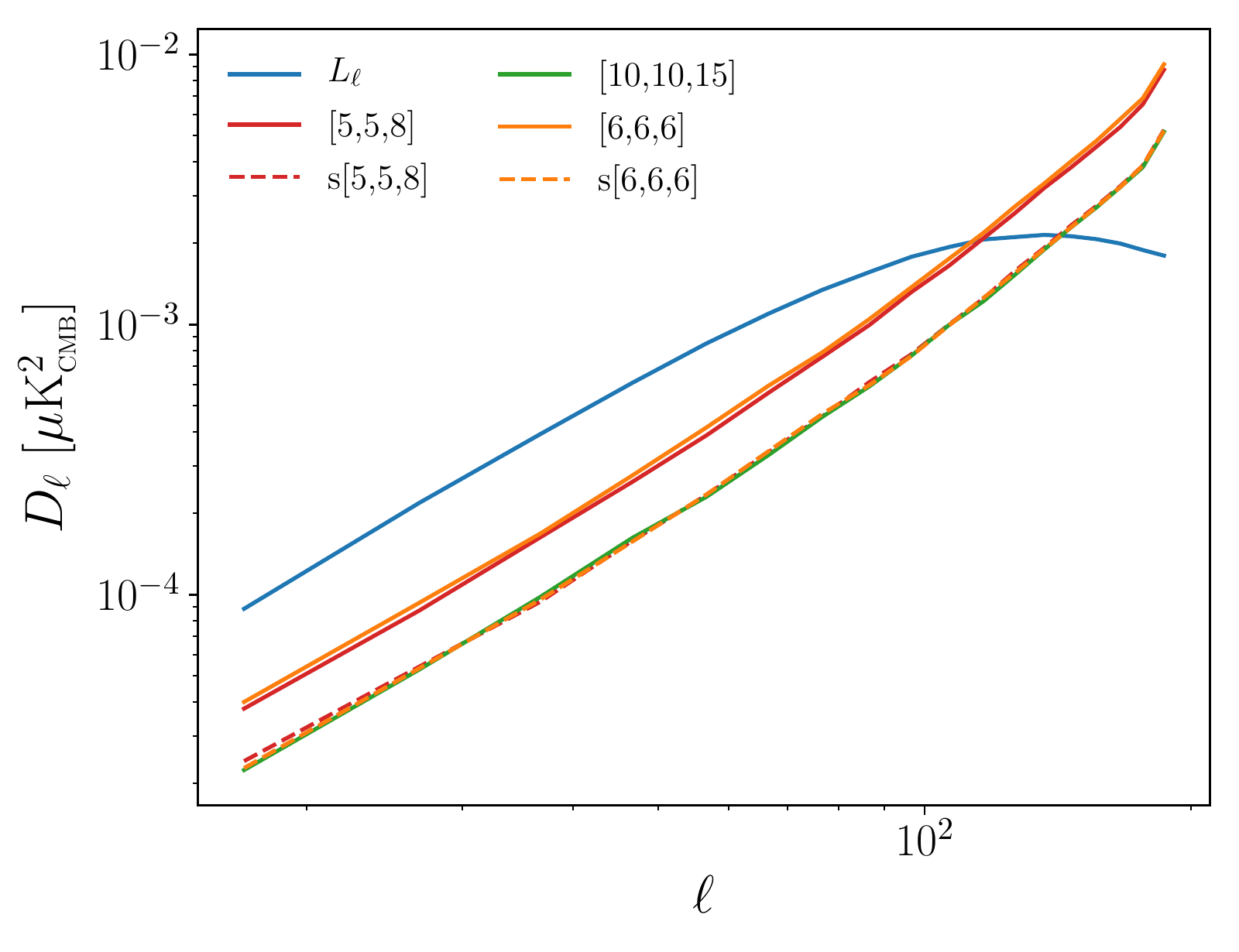}
    \caption{Residuals model power spectra for different telescope setups. The dashed (solid) lines correspond to a scenario where the instrument noise is (not) scaled.}
    \label{fig:residuals_setups_comparison}
\end{figure}

\begin{figure}
    \begin{subfigure}{.31\linewidth}
        \includegraphics[width=\linewidth]{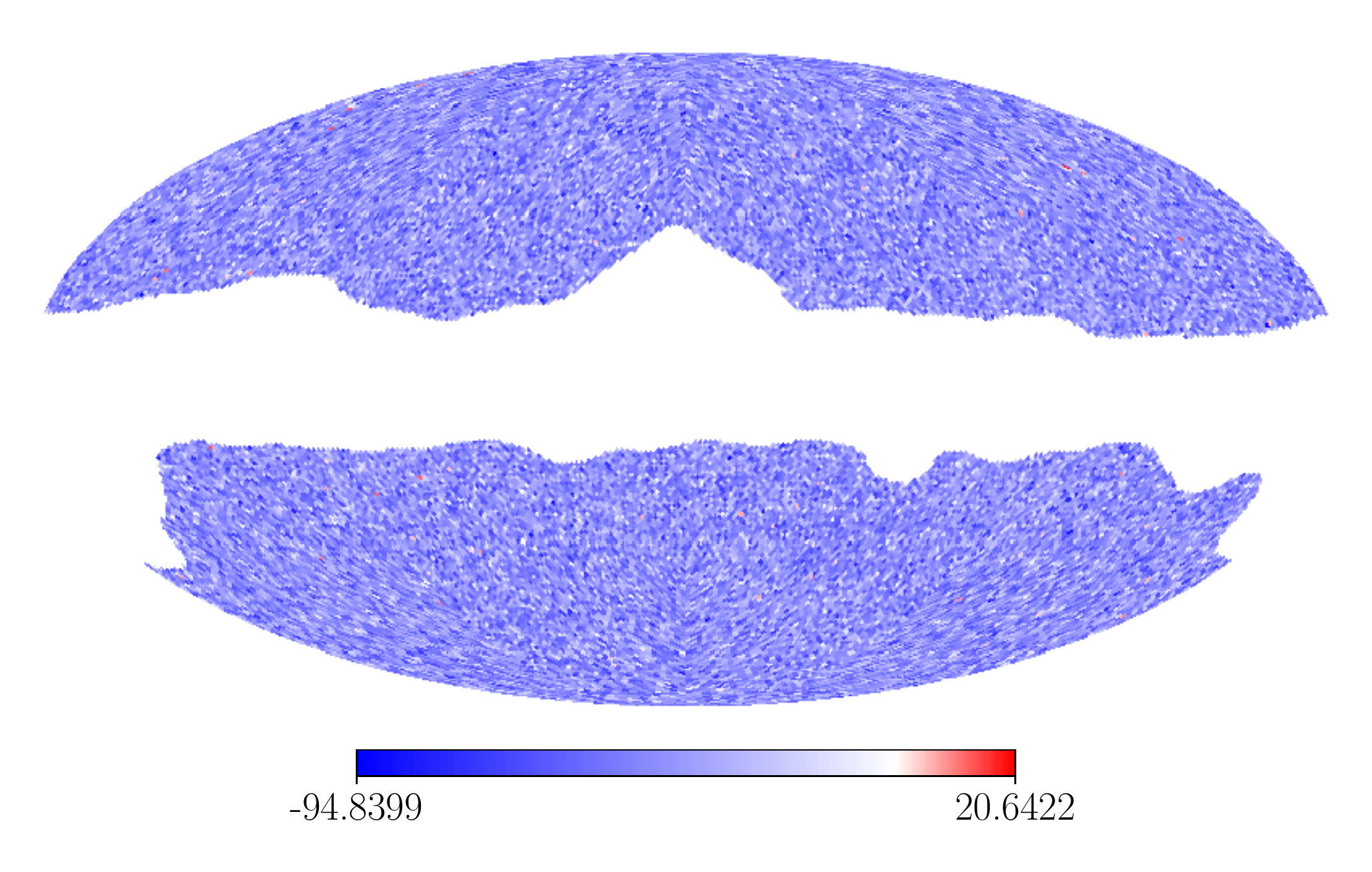}
        \caption{Default - 558-scaled}
        \label{subfig:BIC_maps_default_558_scaled}
    \end{subfigure}
        \begin{subfigure}{.31\linewidth}
        \includegraphics[width=\linewidth]{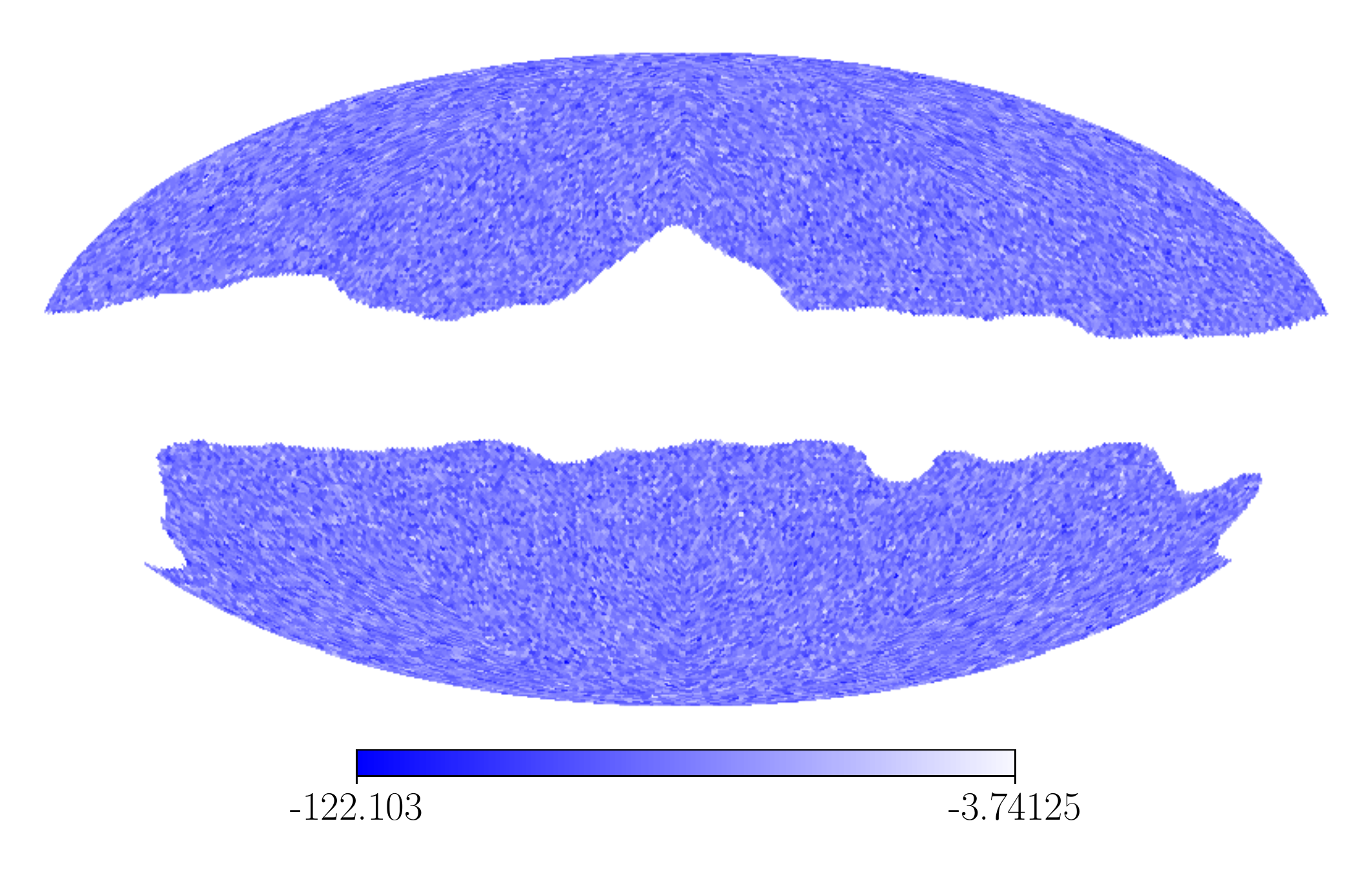}
        \caption{Default - 666-scaled}
        \label{subfig:BIC_maps_default_666_scaled}
    \end{subfigure}
        \begin{subfigure}{.31\linewidth}
        \includegraphics[width=\linewidth]{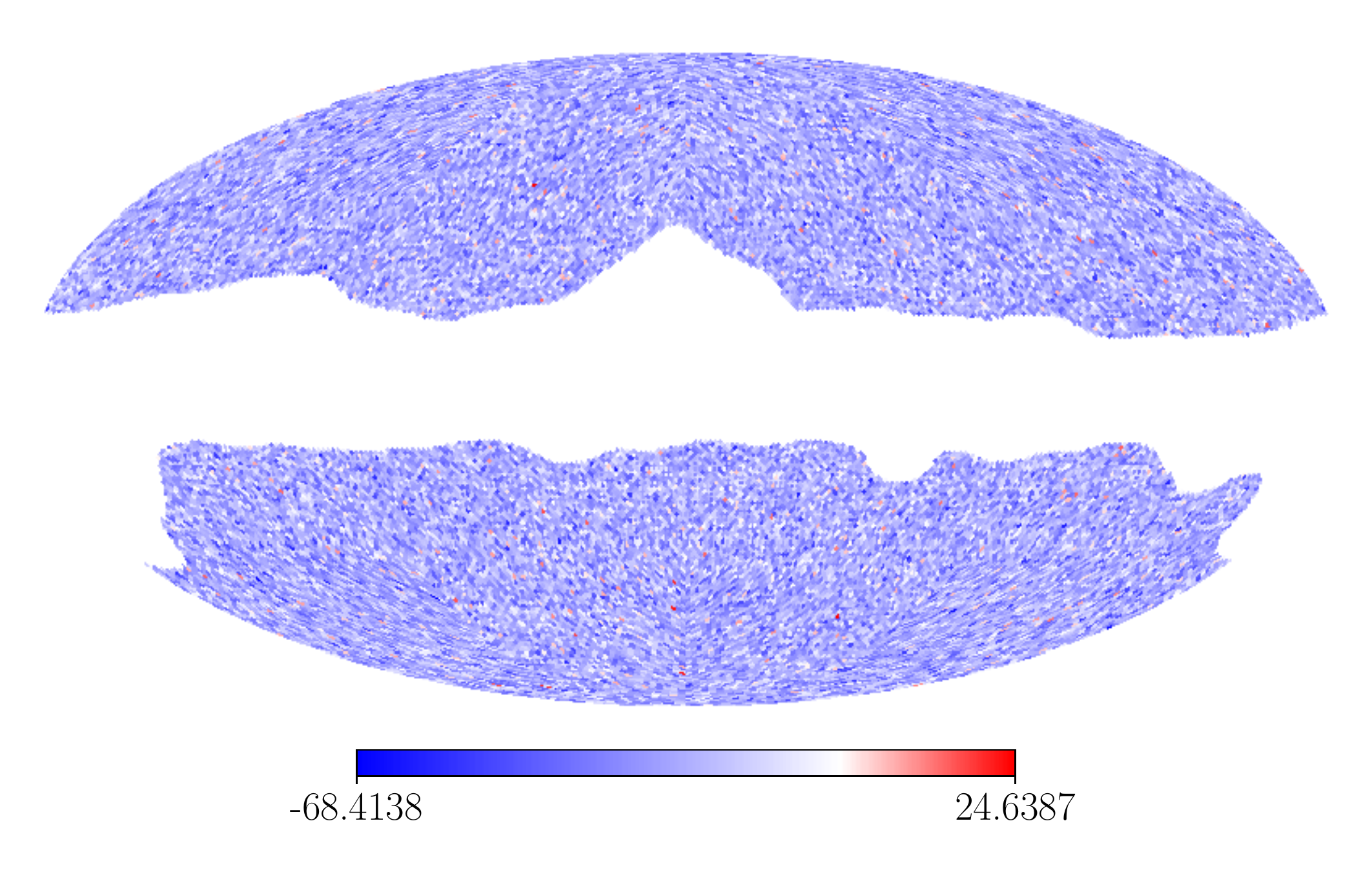}
        \caption{558-scaled - 666-scaled}
        \label{subfig:BIC_maps_558_scaled_666_scaled}
    \end{subfigure}
    \caption{Difference BIC maps.}
    \label{fig:BIC_maps}
\end{figure}

Here, we have compared the instruments performance for three different telescope setups ([10,10,15], [5,5,8] and [6,6,6]) with both, their respective and scaled noise. The results from each setup are shown in table~\ref{tab:scenario_results}.
\begin{table}
    \centering
    \begin{tabular}{|cccccc|}
    \hline
       Scenario &  $\sigma_r \times 10^3$ &  $a_L$ & $\sigma_{a_L}$ & $a_R$ & $\sigma_{a_R}$ \\
    \hline
        default              &  1.0   & 1.01 & 0.05 & 0.99 & 0.04 \\
        558                  &  1.2   & 1.02 & 0.06 & 0.98 & 0.03 \\
        558-scaled           &  1.0   & 0.91 & 0.05 & 1.07 & 0.04 \\
        666                  &  1.2   & 0.92 & 0.07 & 1.04 & 0.03 \\
        666-scaled           &  1.0   & 0.96 & 0.05 & 1.04 & 0.05 \\
        default-delensed     &  0.7   & 0.53 & 0.04 & 0.94 & 0.04 \\
        starobinsky          &  1.1   & 0.92 & 0.05 & 1.06 & 0.05 \\
        starobinsky-delensed &  0.8   & 0.50 & 0.04 & 0.99 & 0.04 \\
        AME                  &  1.1   & 1.00 & 0.06 & 1.00 & 0.04 \\
        NH                   &  1.5   & 0.99 & 0.08 & 1.01 & 0.07 \\
        cosmoareas           &  1.4   & 1.00 & 0.03 & 1.00 & 0.01 \\
        correlated-noise     &  1.0   & 0.99 & 0.05 & 1.02 & 0.04 \\
    \hline
    \end{tabular}
    \caption{\textbf{Fit Results.} The tensor-to-scalar ratio uncertainty $\sigma_r$, as well as the $a_L$ and $a_R$ values and uncertainties obtained from the power spectrum fit for each scenario studied.} 
    \label{tab:scenario_results}
\end{table}\\
\\
 With regard to the $r$ uncertainty, we observe a slight decrease when the noise is scaled. This is a result of the lensing being the principal error source as it is shown in figure~\ref{fig:residuals_setups_comparison}. As a consequence, there is only an improvement at small scales, where the residuals are larger than the lensing.   
\\
\\
In figure~\ref{fig:residuals_setups_comparison}, the residuals model for each configuration is represented. When the noise is scaled the amount of residuals left is the same for all setups. In order to study if the distribution of frequency channels matters when the noise is scaled, we have applied a model selection prior independent criterion that takes into account the number of data points, the Bayesian Information Criteria (BIC) \cite{schwarz1978estimating}:
\begin{equation}
    \textrm{BIC} = -2 \log \mathcal{L} + p\log n_T \, ,
    \label{eq:BIC}
\end{equation}
where $p$ is the number of model parameters. The smaller the BIC score, the better the fit. In figure~\ref{fig:BIC_maps} the difference BIC maps  pair combination of the default, 558-scaled and, 666-scaled are displayed. It is clear that the default setup provides a superior fit than both the 558-scaled and, the 666-scaled. We infer that, given this instrumental noise, the best results are obtained when the number of channels is the largest, i.e., it is better to have more noisier signal channels than fewer more precise channels. 
\begin{figure}
    \centering
    \includegraphics[width=.6\linewidth]{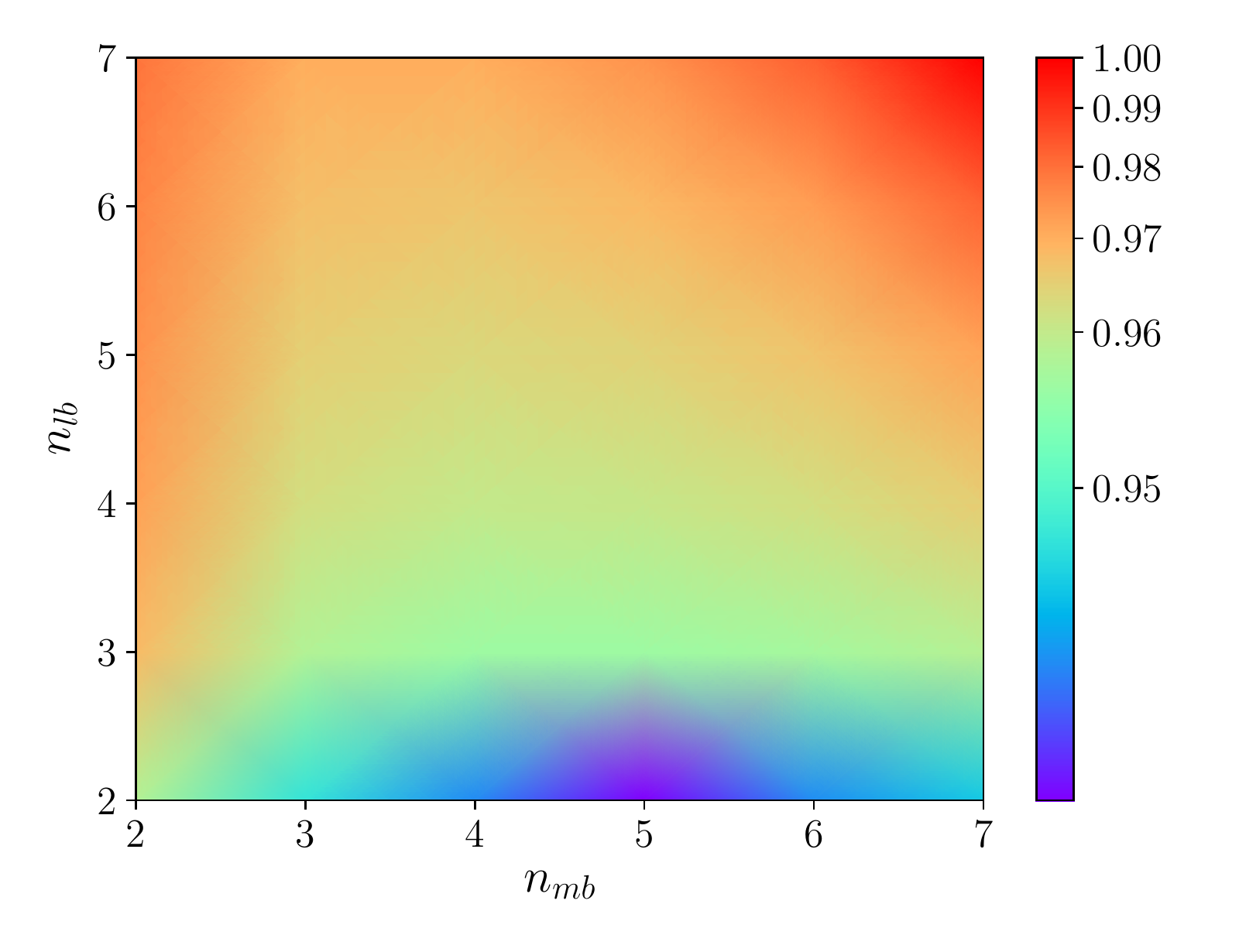}
    \caption{Fisher CMB uncertainty estimation for a typical pixel ($Q$ Stokes parameter) as a function of the $n_{lb}$ and $n_{mb}$. Results are normalized with respect to the maximum value. The number of channels at the $hb$ is $n_{hb} = 18- n_{lb}-n_{mb}$.}
    \label{fig:fisher_cmb_distribution_channels_per_band}
\end{figure}\\
\\
Moreover, the setup 558-scaled yields better results compared to the 666-scaled. To understand these results we have performed a simple study of the dependence of the CMB uncertainty estimation on the distribution of frequency channels among the available bands, given a fixed total number of channels. The CMB uncertainty is estimated using the Fisher matrix obtained from \eqref{eq:likelihood}. As an illustration, in figure~\ref{fig:fisher_cmb_distribution_channels_per_band} the CMB uncertainty for a typical pixel is shown as a function of $n_{lb}$ and $n_{mb}$ in a telescope with $n_T$ fixed to 18 channels and the instrumental noise scaled. We have verified that the behavior observed in the figure is independent of the particular pixel under study and, also the same for both $Q$ and $U$ Stokes parameters. From figure~\ref{fig:fisher_cmb_distribution_channels_per_band} it is inferred that the uncertainty on the CMB parameter is smaller in the 558-scaled scenario than in the 666-scaled, which can explain the results from figure~\ref{subfig:BIC_maps_558_scaled_666_scaled}. Besides, according to this Fisher analysis, the best channel distribution with a fixed total number of channels is a distribution with a few channels at the $lb$, an optimal value at the $mb$ (which in the particular case of $n_T = 18$ is five) and, most channels at the $hb$. The worst is obtained when the $n_{hb}$ is the lowest. Thus, information from the higher frequencies is crucial in this analysis, since those are where the dust information is comprised.

\subsection{Primordial \textit{B}-modes}
\label{subsec:primordial_b_modes}

As previously mentioned, the success of an experiment relies on its ability to constrain $r$, since it has no lower limit. However, some of the preferred theoretical models predict $r$ values that will be either detected or rejected with the target uncertainty ($\sigma_r \simeq 10^{-3}$). In this section we have studied if $r$ is detectable considering the Starobinsky model. Moreover, we have studied the effect of applying a delensing of 50\%, by simulating CMB maps with half the lensing power for $r=0$ and $r =3.7\times 10^{-3}$ (Starobinsky's). The results for each scenario are shown in table~\ref{tab:scenario_results}. \\
\\
First, we observe that within the Starobinsky scenario, $r$ is detectable with more than 3-$\sigma$ even when no delensing is performed as was forecasted in figure~\ref{fig:sigma_r_density_plot_default}. Moreover, when the default and Starobinsky scenarios are compared with their delensed version, we now observe a reduction of the uncertainty. This implies that in order to obtain a stringent constraint on $r$ some sort of delensing will be required.

\subsection{Foreground Model}
\label{subsec:foreground_model}

\begin{figure}
    \centering
    \begin{minipage}[t]{.48\linewidth}
        \centering
        \includegraphics[width=\linewidth]{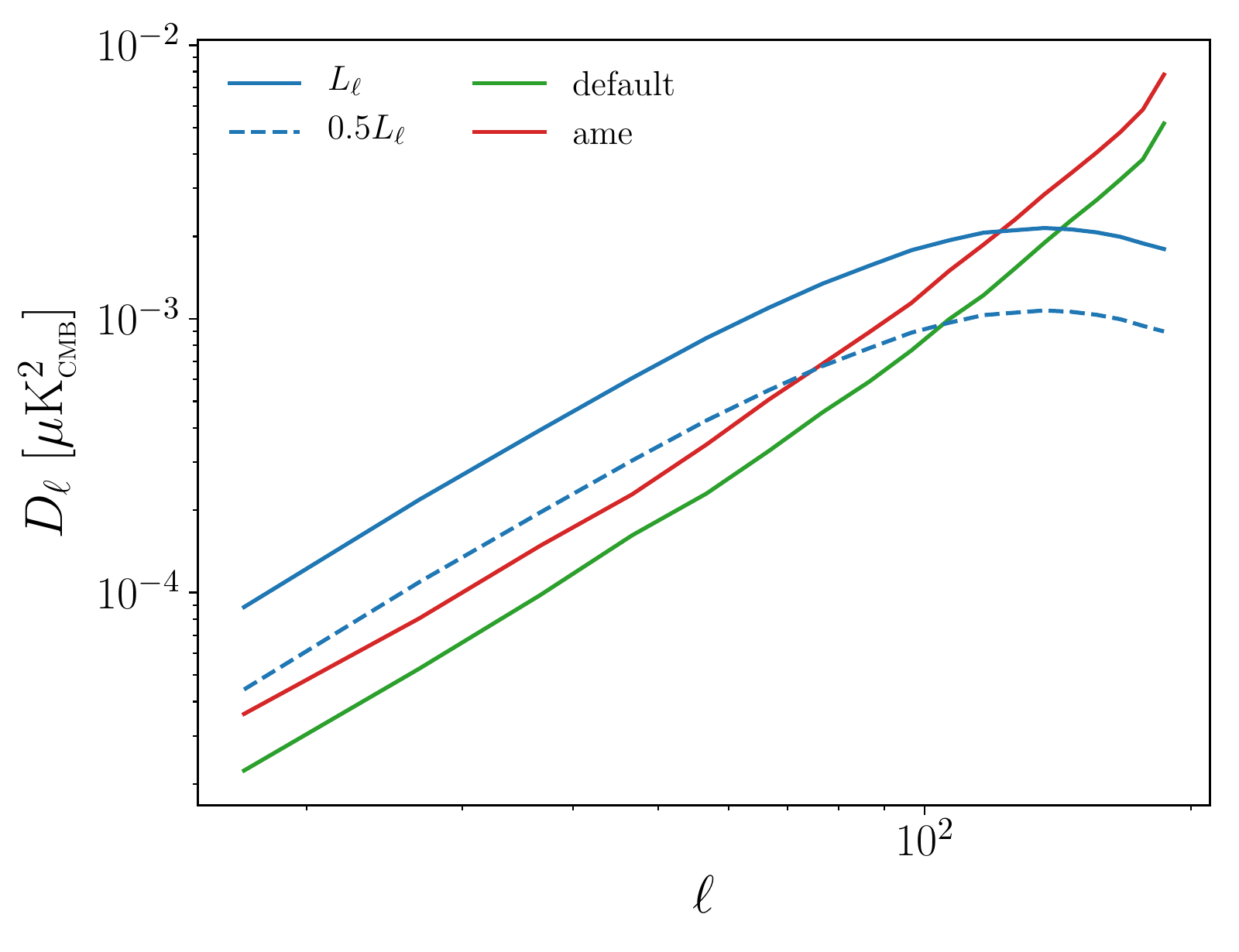}
        \captionof{figure}{Residuals model power spectra for the default and AME scenario compared to the modeled lensing power spectrum when 0 and 50\% delensing is performed.}
        \label{fig:comparison_residuals_sd_sda}
    \end{minipage}
    \hspace{.01\linewidth}
    \begin{minipage}[t]{.48\linewidth}
        \centering
        \includegraphics[width=\linewidth]{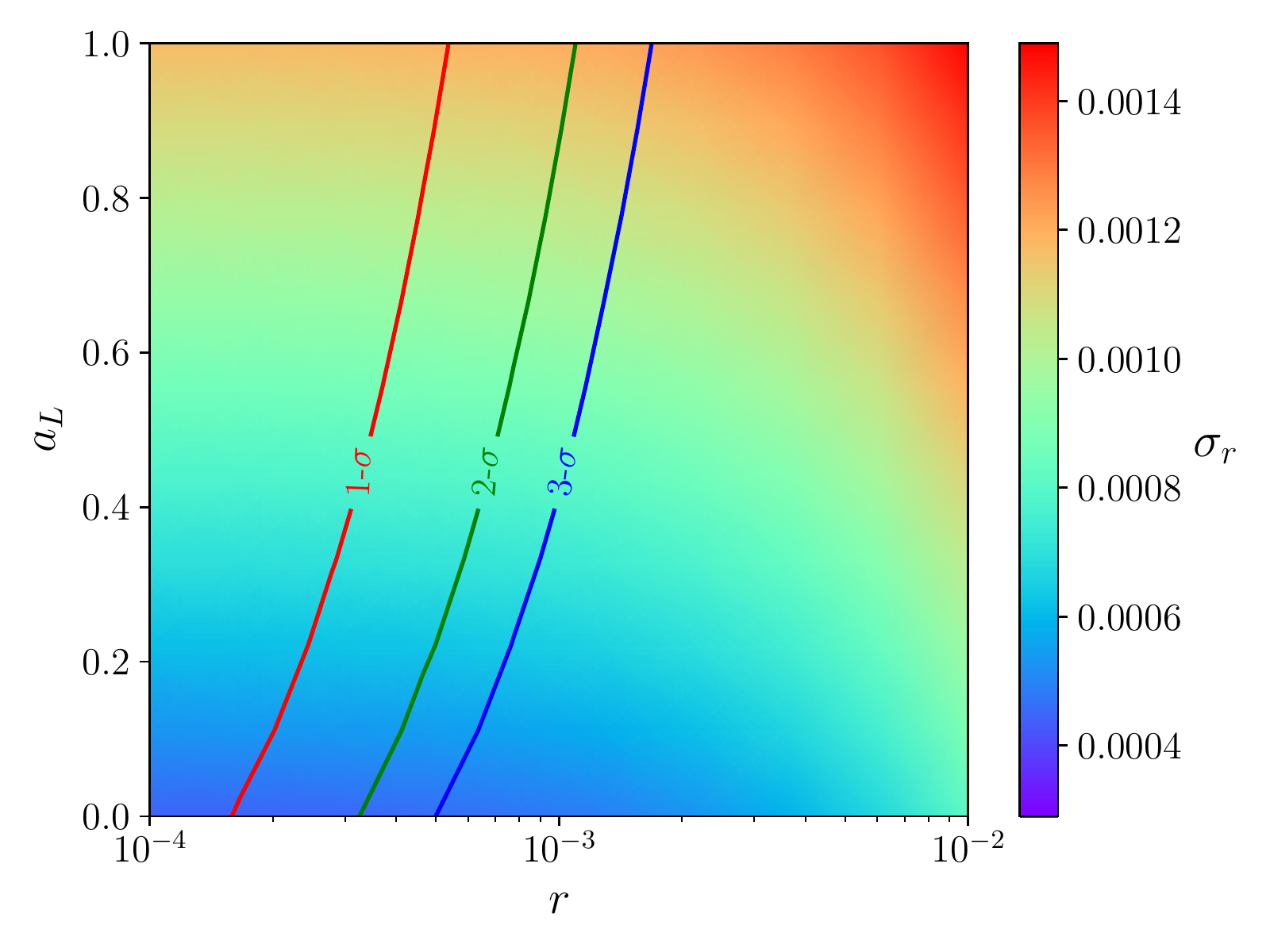}
        \captionof{figure}{Density plot of $\sigma_r$ as a function of $r$ and $a_L$, for fixed $a_R=1$ with the AME's residuals model. The 1,2,3-$\sigma$ signal-to-noise contours are also represented.}
        \label{fig:sigma_r_density_plot_ame}
    \end{minipage}
\end{figure}

In this section, we have explored the possibility of a polarized AME emission and, its implications for the $r$ detection. The results from the fit with this model are also displayed in table~\ref{tab:scenario_results}.\\
\\
The uncertainty on $r$ is slightly higher than in the default scenario mainly due to the increase in the number of model parameters. This is shown in figure~\ref{fig:comparison_residuals_sd_sda}, where the AME's residuals model is proven to be larger than the default's. Moreover, we see that the lensing continues to be the main contaminant, which yields a similar uncertainty in both scenarios. Only if a significant delensing is performed, e.g., reducing half or more the lensing contribution, we observe distinguishable differences between the two models.
\\
\\
The latter argument is confirmed in figure~\ref{fig:sigma_r_density_plot_ame} where in the AME scenario $\sigma_r$ is shown as a function of  $r$ and $a_L$ with fixed $a_R =1$. The results at large $a_L$ are virtually unchanged from the results of  figure~\ref{fig:sigma_r_density_plot_default}, and shifts appear at low $a_L$. Despite this change, $r$ values similar to Starobinsky's can still be detected if a possible AME polarization is taken into account.

\subsection{Observational Strategies}
\label{subsec:observation_strategies}

In this section we compare the results obtained with the default, NH and, cosmoareas scenarios which correspond to three different observational strategies: i) Two experiments located at each hemisphere covering the full sky (default), ii) only one experiment located at the NH (NH), iii) an experiment at the NH exploring small sky patches (cosmoareas). In other words, we are  studying whether a ground-based experiment can reach these scientific goals with just one instrument. An instrument can take measurements on its whole available sky view or, focus on the cleanest areas to achieve better sensitivities. Given the same observational time, the  sensitivities of two different observational strategies (1) and (2) of a single telescope are related by  
\begin{equation}
    s^{(1)} = \sqrt{\dfrac{f^{(1)}_{sky}}{f^{(2)}_{sky}}}s^{(2)} \, ,
    \label{eq:sensitivity_fsky}
\end{equation}
where $f_{sky}^{(l)}$ is the fraction of the sky covered by the strategy $(l)$.\\
\\
Therefore, the LFS's sensitivity in the cosmoareas scenario is corrected by the factor with respect to the NH scenario defined in \eqref{eq:sensitivity_fsky}. The fraction of available observable sky is $f_{sky}^{(ca)} = 0.08$ ($f_{sky}^{(NH)} = 0.42$)  in the cosmoareas (NH) scenario\footnote{Notice that the $f_{sky}$ considered in the NH scenario is the fraction of observable sky in the \texttt{QUIJOTE} Wide Survey mask, i.e., the galactic mask used in the analysis is not taken into account.}. 
\begin{figure}
    \centering
    \begin{subfigure}{.48\linewidth}
        \centering
        \includegraphics[width=\linewidth]{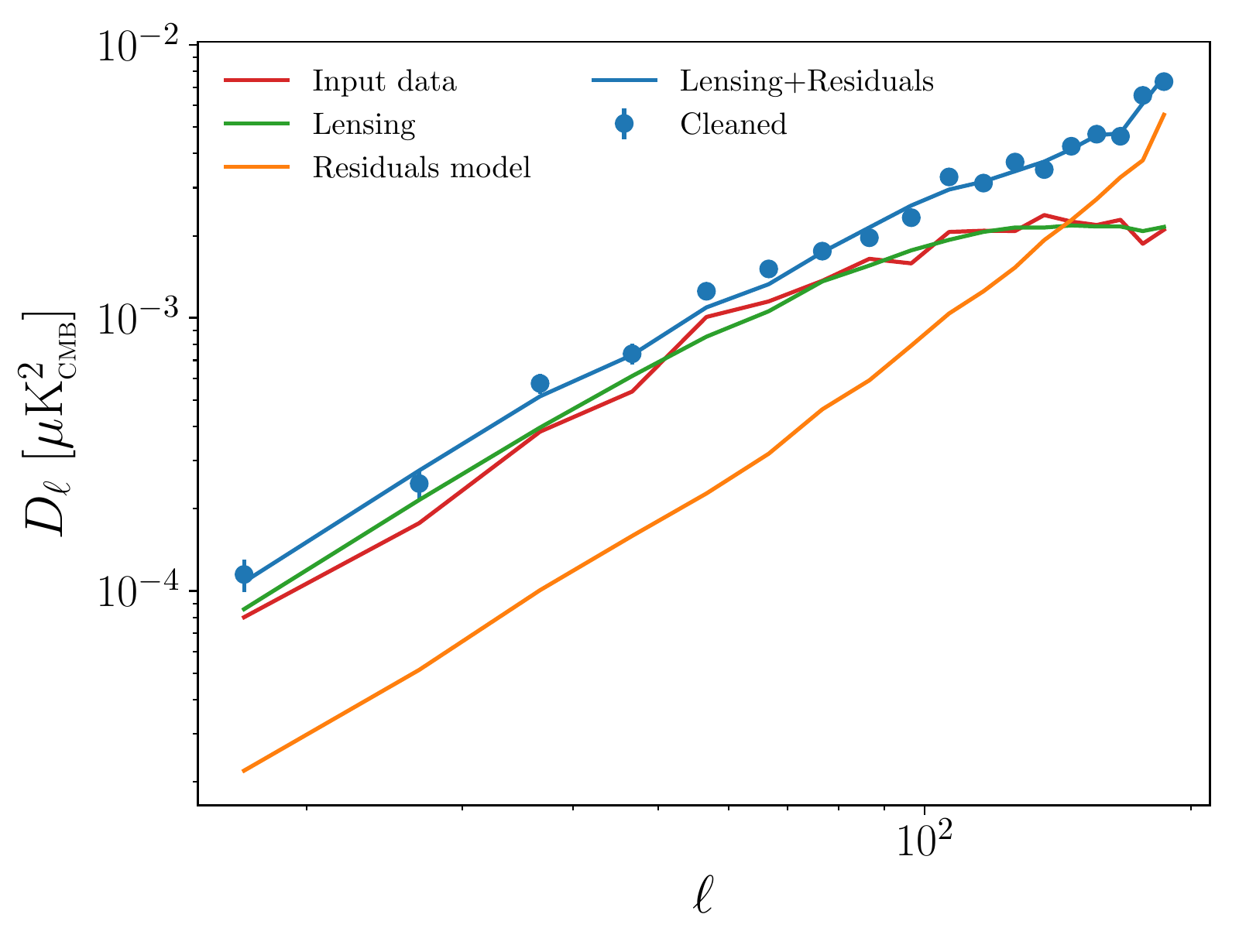}
        \caption{NH}
        \label{subfig:plot_results_NH}
    \end{subfigure}
    \hspace{.01\linewidth}
    \begin{subfigure}{.48\linewidth}
        \centering
        \includegraphics[width=\linewidth]{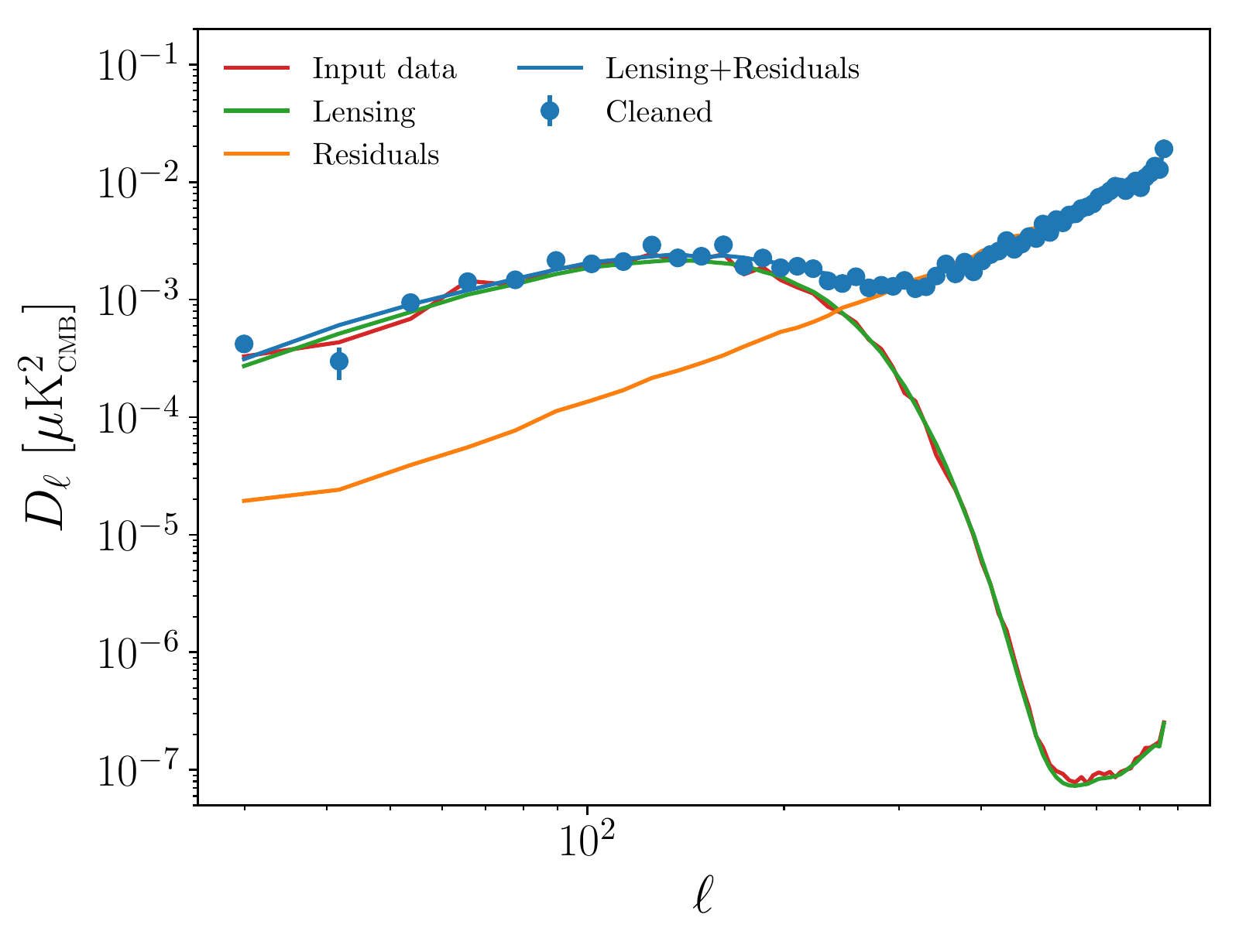}
        \caption{cosmoareas}
        \label{subfig:plot_results_cosmoareas}
    \end{subfigure}
    \caption{CMB $B$-mode power spectra contributions in a given scenario when the sensitivity accounts for the observed sky size.}
    \label{fig:plot_results_NH_CA}
\end{figure}\\
\\
The fit results are shown in table~\ref{tab:scenario_results}. As expected the default scenario has a smaller $r$ uncertainty compared to the NH scenario since it covers a wider sky area. On the other hand, if measurements are restricted to a particular hemisphere, we observe that it is better to cover smaller regions than the whole available sky, not only due to the smaller $\sigma_r$ obtained but notably because the residuals foregrounds and lensing contributions are better characterized. \\ 
\\
Figure~\ref{subfig:plot_results_NH} and figure~\ref{subfig:plot_results_cosmoareas} depict the power spectra results in the NH and cosmoareas scenarios respectively. We observe from the power spectra comparison of the cosmoareas scenario with the default (figure~\ref{fig:plot_results_default}), that in the former strategy the residuals decrease at the recombination bump. 

\subsection{Atmospheric/Systematics Contamination}
\label{subsec:atmospheric_contamination}

\begin{figure}
    \centering
    \begin{minipage}{.48\linewidth}
        \centering
        \includegraphics[width=\linewidth]{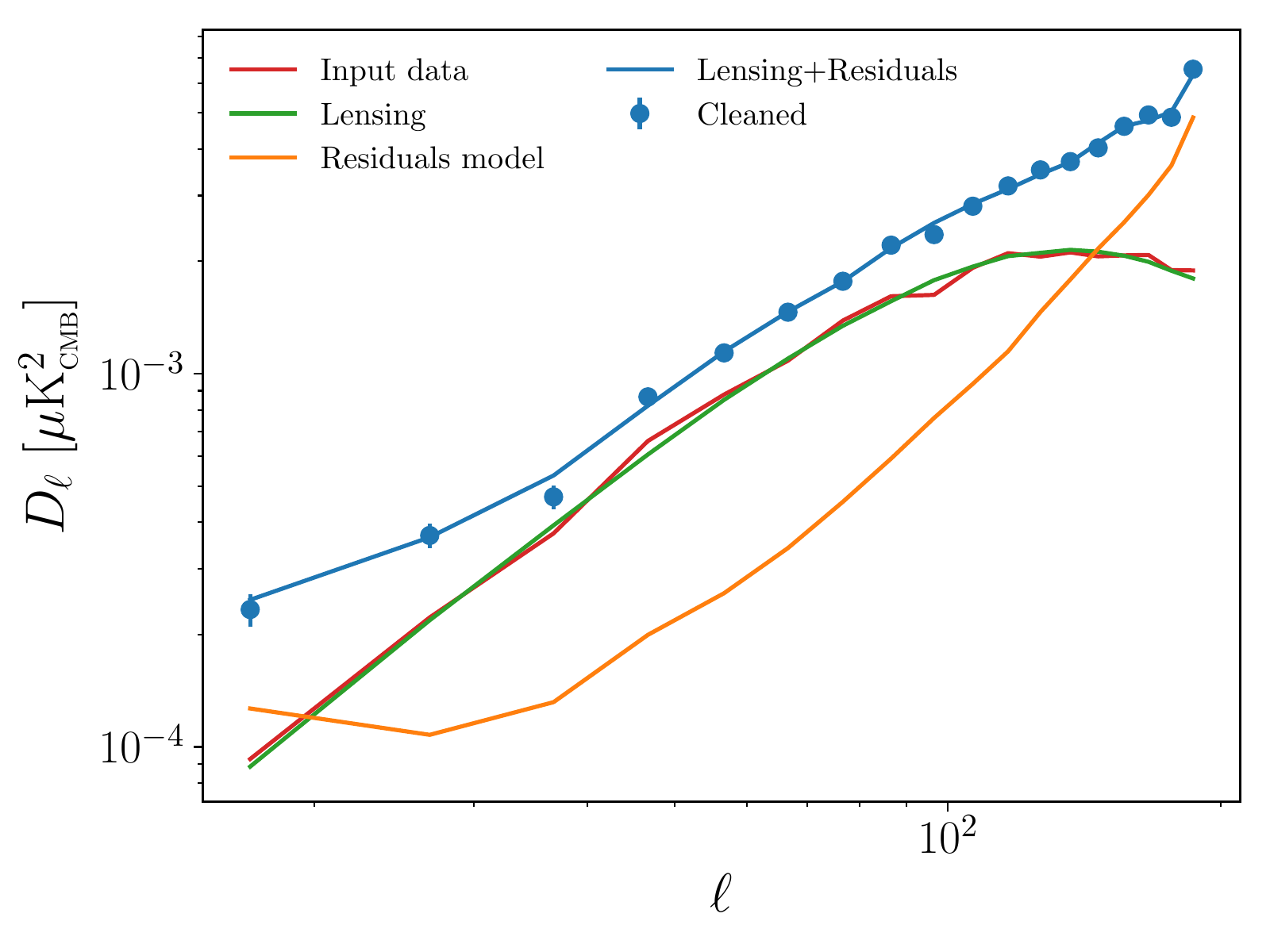}
        \captionof{figure}{CMB $B$-mode power spectra contributions in the correlated-noise scenario.}
        \label{fig:plot_results_atmosphere}
    \end{minipage}
        \hspace{.01\linewidth}
    \begin{minipage}{.48\linewidth}
        \footnotesize
        \centering
        \begin{tabular}{|cc|}
            \hline
            Band [GHz] & s[$\mu$K arcmin] \\
            \hline
            40 & 37.5 \\
            50 & 24.0 \\
            60 & 19.9 \\
            68 & 16.2 \\
            78 & 13.5 \\
            89 & 11.7 \\
            100 & 9.2 \\
            119 & 7.6 \\
            140 & 5.9 \\
            166 & 6.5 \\
            195 & 5.8 \\
            235 & 7.7 \\
            280 & 13.2 \\
            337 & 19.5 \\
            402 & 37.5 \\
            \hline
        \end{tabular}
        \captionof{table}{\textbf{LiteBIRD's characteristics.} Channels' central frequency and sensitivity \cite{hazumi2019litebird}.}
        \label{tab:LiteBIRD_setup}
    \end{minipage}
\end{figure}

If atmospheric contamination in polarization is sufficiently large, its  impact can have detrimental consequences on the uncertainty of $r$ as it was shown in  \cite{alonso2017simulated}. Here, we study the repercussions of including a correlated noise that mimics the atmosphere and/or systematics contribution. The results of the correlated-noise scenario is shown in table~\ref{tab:scenario_results}.  \\
\\
From the results, we see that the uncertainty on $r$ does not increase. Comparing the residuals model in the default and the atmosphere scenario we see that the increment only occurs at low $\ell$ where the residuals are at their lowest. Thus, the residuals contribution to the uncertainty comes primarily from the small scales where both scenarios have similar amount of residuals.   \\
\\
From figure~\ref{fig:plot_results_atmosphere}, we observe that our methodology is able to recover the correlated noise introduced by the atmosphere at large scales. This yields a reduction on possible biases on $r$ due to incorrect noise removal. On the other hand, if atmospheric noise is not negligible at the experiment's location site, the residuals increment at low multipoles, then it prevents the information gathering from the first bump even if the available observed sky is sufficiently large to measure it. 

\section{Complementarity to LiteBIRD}
\label{sec:complementarity_LiteBIRD}

LiteBIRD, Lite (Light) satellite for the studies of $B$-mode polarization and Inflation from cosmic background Radiation Detection, is a planned satellite mission whose primary objective is the detection of PGWs through the footprint left on the polarized CMB $B$-modes \cite{hazumi2019litebird}. Recently, it has been selected as the Japanese Space Agency's (JAXA’s) strategic large mission and its launch is planed for 2027. Its design is optimized for CMB $B$-mode detection on large angular scales, and its principal scientific goal is reaching $\sigma_r (r=0) \leq 10^{-3}$. \\
\\
LiteBIRD's frequency coverage ranges from 40 to 402 GHz, as can be seen in  table~\ref{tab:LiteBIRD_setup}, where LiteBIRD's observing frequency bands and sensitivities are shown. If AME has a polarized contribution, its detection by LiteBIRD would be extremely challenging (see figure~\ref{fig:ame_fit}), which may result in a potential bias for the primordial $B$-mode detection. A joint analysis with LFS could enhance LiteBIRD's capabilities for the low-frequency foregrounds characterization. Even if the AME does not have a measurable polarized signal, LiteBIRD can benefit significantly from a joint analysis as our experiment covers smaller frequencies where the synchrotron is considerably larger, see figure~\ref{fig:foregrounds_intensity}.
\\
\\
Below we analyze the improvement on the foreground characterization when both experiments are combined.

\paragraph{Foregrounds Characterization.}

\begin{figure}
    \centering
    \begin{subfigure}{0.76\textwidth}
        \centering
        \includegraphics[width=\linewidth]{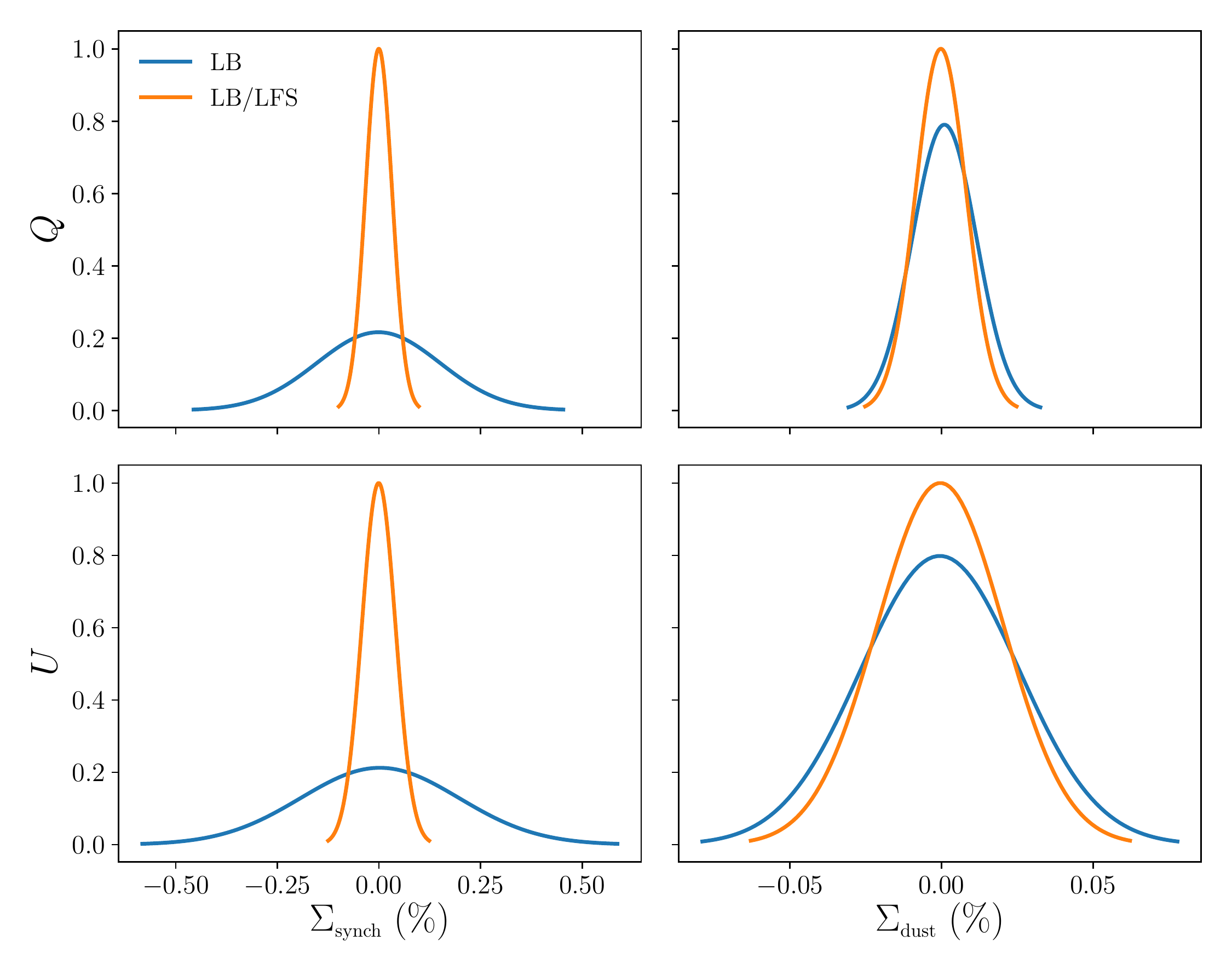}
        \caption{Synchrotron and dust relative residuals at 70 GHz.}
        \label{subfig:foregrounds_histograms_synchdust}
    \end{subfigure}
    \begin{subfigure}{0.76\textwidth}
        \centering
        \includegraphics[width=\linewidth]{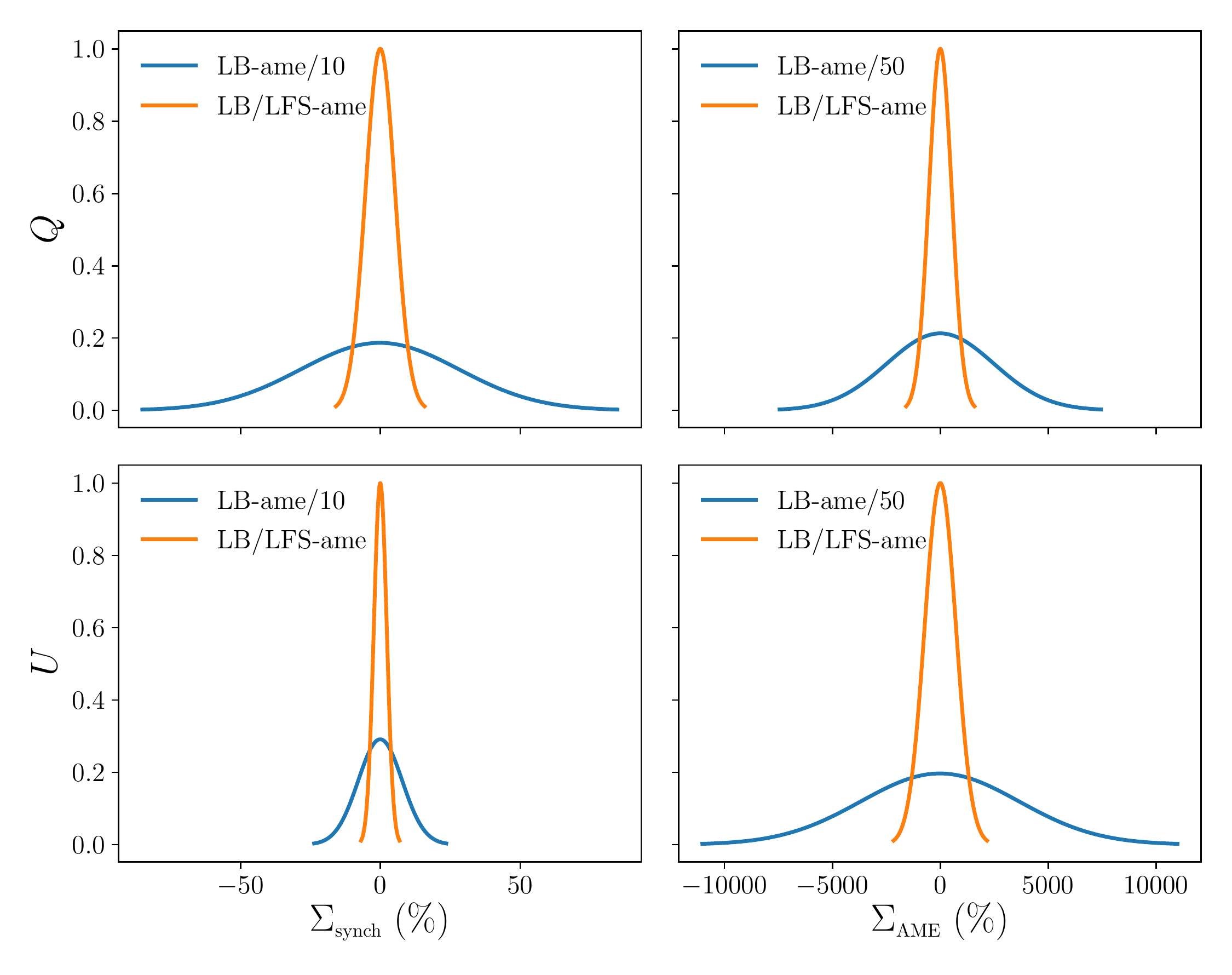}
        \caption{Synchrotron and AME relative residuals at 70 and 23 GHz respectively.}
        \label{subfig:foregrounds_histograms_synchame}
    \end{subfigure}
    \caption{Foregrounds relative residuals distributions in $Q$ (top) and $U$ (bottom) Stokes parameters for LiteBIRD alone (blue), and LiteBIRD combined with this instrument (orange). The distributions are normalized with respect to the the LB/LFS distribution's maximum.}
    \label{fig:foregrounds_histograms}
\end{figure}
We have studied an experiment's ability to characterize foregrounds by studying the foregrounds residuals relative distributions at a given frequency, i.e., the ratio of the difference between the input modeled foregrounds and the retrieved ones over the input modeled foregrounds, using the corresponding models given by equations \eqref{eq:synchrotron_model}-\eqref{eq:ame_model}. In figure~\ref{fig:foregrounds_histograms}, the LB (LB-AME) and LB/LFS (LB/LFS-AME) scenarios' foregrounds residuals distributions are compared. Notice from table~\ref{tab:scenarios} that the LFS considered here has only frequency channels at the lb and mb, since the frequencies at hb overlap with LiteBIRD's frequency range.  \\
\\ 
In figure \ref{subfig:foregrounds_histograms_synchdust} the synchrotron's and dust's residuals distributions are compared for the LB and LB/LFS scenarios at 70 GHz. We observe no significant difference in the dust residuals when LiteBIRD is combined with LFS, which is indeed expected as the latter focus mainly on low-frequencies where the thermal dust is insignificant. On the other hand, the inclusion of the ground-based experiment narrows considerably the synchrotron's residuals distribution. Therefore, a joint analysis will reduce the uncertainties relative to the component separation. Moreover, if the actual synchrotron model is more complex than current models, more channels at the low-frequency spectrum regime might become essential to reconstruct properly the CMB. \\
\\
Figure \ref{subfig:foregrounds_histograms_synchame} compares the synchrotron's and AME's residuals distributions at 70 and 23 GHz respectively for the LB-AME and, LB/LFS-AME scenarios. Note that the synchrotron (AME) distributions in the LB-AME case are divided by a factor of 10 (50) for a better visualization. As in figure \ref{subfig:foregrounds_histograms_synchdust} the synchrotron is better recovered when both experiments are combined. However, the most outstanding improvement is observed in the AME's residuals distribution. LiteBIRD's frequency range does not overlap with the range where the AME is most dominant. If AME happens to be slightly polarized, a bias could be introduced in the CMB during the component separation. A  combined analysis will help overcome this issue.
\\
\\
If current foreground modelling is a sufficiently good approximation of the diffuse sky, LiteBIRD can benefit from a joint analysis by reducing the uncertainties in the foregrounds recovery. Moreover, if current modelling lacks the low-frequency foregrounds' inherent complexity, LiteBIRD could always employ data from low-frequency experiments such as the one proposed here.

\section{Conclusions}
\label{sec:conclusions}

We have studied whether a ground-based telescope operating in the low-frequency regime is able to detect or, at least, constrain $r$ at the level of $\sigma_r(r=0) = 10^{-3}$. For this purpose, we have applied a full-parametric pixel-based maximum likelihood component separation method to obtain the CMB as well as the foregrounds parameters. Moreover, we developed a self-consistent approach to estimate the residuals left from the component separation methodology. Finally, with these techniques we have tested different scenarios that an experiment of this sort can face. \\
\\
First of all, we have compared different LFS's frequency channels distributions with diverse telescope sensitivities, to obtain the most optimal telescope configuration. We have seen that given the same effective telescope sensitivity, it is preferable to have more noisier channels than a few channels with large sensitivities. Besides, it has been shown that the channels at the high frequency band help trace the thermal dust information.\\
\\
We have found that $r$ values within the Starobinsky's range are detectable with this type of experiment even if no delensing is performed. However, some sort of delensing should be performed in order to reduce the $r$ uncertainty, $\sigma_r$. This comes as a result of the lensing being the principal $BB$ power spectrum  contaminant in this experimental configuration. \\
\\
Since the foreground sky could be more complex than what  current models predict, we have also studied the LFS's performance when a polarized AME contribution is included. The results are virtually the same when no delensing is applied, as it constitutes the primary source of error. However, when a large fraction of lensing is removed, we estimated from the residuals left in the model with AME that the region where $r$ is no longer detectable is larger than when AME is not taken into account. \\
\\
In this study we have also considered different observational strategies related to the number of available telescope's locations. For a experiment located at a given hemisphere, focusing on small sky patches yields smaller residuals at the small scales, and an overall smaller $\sigma_r$. However, we shown that the tightest constraints on $r$ are obtained when the largest amount of sky is covered, i.e., at least one instrument per hemisphere.\\
\\
Furthermore, we have considered the contamination by the atmosphere and/or systematics in polarization measurements. These contributions introduce a correlated noise that masks the power spectrum at low multipoles, i.e., at large scales. With our residuals estimation method, we are able to recover the uncertainty generated by the correlated noise and avoid biases in the fit. \\
\\
Finally, we have studied its potential complementarity with LiteBIRD. The LFS experiment explores the low-frequency range with a sensitivity never achieved before. Therefore, it significantly improves the characterization of the low-frequency foregrounds. Moreover, the need for a joint analysis will be more justified if the low-frequency sky is more complex than expected nowadays.\\
\\
In conclusion, this type of experiment alone is capable of detecting $r$ (in the Starobinsky model), or at least constrain it with $\sigma_r= 10^{-3}$ even when no delensing is performed. Additionally, it will help with the low-frequency foregrounds characterization as it reaches unprecedented sensitivities in this regime. The latter makes this instrument also a valuable complement to other satellite and on-ground experiments.

\acknowledgments

 EdlH acknowledge partial financial support from the \textit{Concepci\'on Arenal Programme} of the Universidad de Cantabria. We acknowledge Santander Supercomputacion support group at the University of Cantabria who provided access to the supercomputer Altamira Supercomputer at the Institute of Physics of Cantabria (IFCA-CSIC), member of the Spanish Supercomputing Network, for performing simulations/analyses. The authors would like to thank Spanish Agencia Estatal de Investigaci\'on (AEI, MICIU) for the financial support provided under the projects with references
ESP2017-83921-C2-1-R and AYA2017-90675-REDC, co-funded with EU FEDER funds, and also acknowledge the funding from Unidad de Excelencia Mar{\'\i}a de Maeztu (MDM-2017-0765).  We make use of the \texttt{HEALPix} \cite{gorski2005healpix},  \texttt{CAMB} \cite{lewis2011camb} and, the  \texttt{PySM}, \texttt{emcee}, \texttt{healpy}, \texttt{NaMaster}, \texttt{matplotlib} and, \texttt{numpy} Python packages.



%
%
\bibliographystyle{unsrt}  
\bibliography{references}

\end{document}